\documentclass[%
 aip, apl, %
 longbibliography, %
 amsmath,amssymb,
reprint
]{revtex4-1}

\usepackage{graphicx}
\usepackage{dcolumn}
\usepackage{bm}
\usepackage[utf8]{inputenc}
\usepackage{makecell} 
\usepackage[T1]{fontenc}
\usepackage{mathptmx}
\usepackage{etoolbox}
\usepackage{float} 
\usepackage{color}
\usepackage{siunitx}
\usepackage{amsmath}
\usepackage[normalem]{ulem}
\usepackage{hyperref}

\makeatletter
\def\@email#1#2{%
 \endgroup
 \patchcmd{\titleblock@produce}
  {\frontmatter@RRAPformat}
  {\frontmatter@RRAPformat{\produce@RRAP{*#1\href{mailto:#2}{#2}}}\frontmatter@RRAPformat}
  {}{}
}%
\makeatother

\usepackage[svgnames]{xcolor}

\begin{document}

\preprint{AIP/123-QED}

\title[Probing Topological Surface States and Conduction via Extended Defects in (Bi$_{1-x}$Sb$_{x}$)$_{2}$Te$_{3}$ Films]{Probing Topological Surface States and Conduction via Extended Defects in (Bi$_{1-x}$Sb$_{x}$)$_{2}$Te$_{3}$ Films}

\author{A. Liu}
\affiliation{Department of Materials Science and Engineering, University of Michigan, Ann Arbor, MI 48109, USA}
\author{A. Gil}
\affiliation{Applied Physics Program, University of Michigan, Ann Arbor, MI 48109, USA}
\author{M. Choi}
\affiliation{Department of Mechanical Science and Engineering, University of Illinois, Champaign, IL 61801, USA}
\affiliation{Materials Research Laboratory, University of Illinois, Urbana, IL, 61801, USA}
\author{B. A. Hanedar}
\affiliation{Department of Physics, Kirklareli University, Kirklareli 39100, Türkiye}
\author{V. Stoica}
\affiliation{Department of Physics, University of Michigan, Ann Arbor, MI 48109, USA}
\affiliation{Department of Materials Science and Engineering, Penn State University, University Park, PA, 16802, USA}
\author{Z. You}
\affiliation{Department of Physics, University of Michigan, Ann Arbor, MI 48109, USA}
\author{S. Sinha}
\affiliation{Applied Physics Program, University of Michigan, Ann Arbor, MI 48109, USA}
\author{T. Hakioglu}
\affiliation{Department of Physics, University of Michigan, Ann Arbor, MI 48109, USA}
\affiliation{Energy Institute, Istanbul Technical University, Maslak, Sarıyer, 34469, Istanbul, Türkiye}
\author{H. T. Johnson}
\affiliation{Department of Mechanical Science and Engineering, University of Illinois, Champaign, IL 61801, USA}
\affiliation{Materials Research Laboratory, University of Illinois, Urbana, IL, 61801, USA}
\author{K. Sun}
\affiliation{Department of Physics, University of Michigan, Ann Arbor, MI 48109, USA}
\author{R. Clarke}
\affiliation{Applied Physics Program, University of Michigan, Ann Arbor, MI 48109, USA}
\affiliation{Department of Physics, University of Michigan, Ann Arbor, MI 48109, USA}
\author{C. Uher}
\affiliation{Department of Physics, University of Michigan, Ann Arbor, MI 48109, USA}
\author{C. Kurdak}
\affiliation{Applied Physics Program, University of Michigan, Ann Arbor, MI 48109, USA}
\affiliation{Department of Physics, University of Michigan, Ann Arbor, MI 48109, USA}
\author{R. S. Goldman}
 \thanks{Corresponding author: rsgold@umich.edu}
\affiliation{Department of Materials Science and Engineering, University of Michigan, Ann Arbor, MI 48109, USA}
\affiliation{Applied Physics Program, University of Michigan, Ann Arbor, MI 48109, USA}
\affiliation{Department of Physics, University of Michigan, Ann Arbor, MI 48109, USA}


\begin{abstract}

(Bi$_{1-x}$Sb$_x$)$_2$Te$_3$ alloys are non-degenerate topological insulators (TIs) whose Dirac point (DP) can be tuned within the bulk bandgap by varying the composition, effectively reducing bulk conduction while allowing surface carrier conduction. Magnetotransport measurements of a series of (Bi$_{1-x}$Sb$_x$)$_2$Te$_3$ thin films indicate electron-dominated conduction, with weak anti-localization attributed to topological surface states (TSSs). Due to the similarity of phase coherence lengths and twin boundary spacings ($\sim$100 nm), we consider the role of twin boundaries as additional conducting paths. Density functional theory calculations reveal an enhanced density of states near the Fermi level at $60^\circ$ twin boundaries, with 2D carrier concentration in excess of $3 \times 10^{13}$ cm$^{-2}$. Furthermore, an analysis of the longitudinal magnetoconductivity yields an upper bound of $7.3 \times 10^{-4}$ S for twin boundary conductivity, resulting in a carrier mobility as high as $142$ cm$^2$/(V$\cdot$s). We discuss the role of twin boundaries in facilitating a transition from a massive Dirac cone dispersion to gapless, topologically protected surface states. Understanding the role of twin boundaries on carrier conduction in non-degenerate TIs is critical for the development of novel TI-based electronic devices.

\end{abstract}

\maketitle

\section{Introduction}

Topological insulators, such as Bi$_2$Se$_3$-, Bi$_2$Te$_3$- and Sb$_2$Te$_3$-based alloys, are an exciting class of quantum materials possessing a bulk band gap with accessible topological surfaces states.\cite{mazumderBriefReviewBi2Se32021, heremansTetradymitesThermoelectricsTopological2017a, tangComprehensiveReviewBi2Te3based2022} In particular, (Bi$_{1-x}$Sb$_x$)$_2$Te$_3$ alloys contain combinations of Te$_{Bi}$, V$_{Sb}$, and Sb$_{Te}$, leading to carrier compensation with both the Fermi level and the DP inside the gap; therefore, these alloys are leading candidates for non-degenerate TI. \cite{zhang_band_2011,chen_experimental_2009,hsieh_observation_2009} Near the Fermi level, the linear dispersion of TSSs gives rise to Dirac fermion behavior,\cite{zhang_topological_2009, xia_observation_2009} enabling spin-momentum locking\cite{hsieh_tunable_2009} without backscattering,\cite{maciejko_quantum_2011} both critical for spintronics and quantum computing.\cite{roushan_topological_2009,zhang_experimental_2009,moore_birth_2010}

During the past decade, TIs have been extensively explored, but the proposed conduction through extended defects has been rarely considered.\cite{ranOnedimensionalTopologicallyProtected2009} For example, in Bi$_2$Se$_3$, scanning tunneling spectroscopy reveals variations in the DP energy across grain boundaries.\cite{liu_tuning_2014} In Bi$_2$Te$_3$, Hall measurements (Van der Pauw configuration) suggest that $60^\circ$ twin boundaries provide bulk carriers.\cite{kimFreeelectronCreation60deg2016} Since Bi$_2$Se$_3$ and Bi$_2$Te$_3$ are degenerate TIs, the contributions of bulk conduction could not be ruled out.

Therefore, in this paper, we use a non-degenerate TI with suppressed bulk conduction to reveal the influence of extended defects on carrier transport in TIs. Using a combination of scanning tunneling spectroscopy, low-field transport measurements, high resolution scanning transmission electron microscopy (STEM), and density functional theory (DFT) calculations, we demonstrate the role of $60^\circ$ twin boundaries in providing additional conduction paths in (Bi$_{1-x}$Sb$_{x}$)$_{2}$Te$_{3}$ thin films, with 2D carrier concentration in excess of $3\times10^{13}\ \rm{cm}^{-2}$ and the mobility as high as 142 cm$^2$/(V$\cdot$s). In addition, the role of twin boundaries in facilitating a transition from a massive Dirac cone dispersion to a gapless, topologically protected surface state is discussed. Identification of conducting pathways associated with extended defects in non-degenerate TIs provides new opportunities for the development of TI-based electronic devices.

\section{Methods}

For these investigations, a series of (Bi$_{1-x}$Sb$_{x}$)$_{2}$Te$_{3}$ films were prepared on sapphire (0001) substrates by molecular-beam epitaxy using $>$ 99.9999$\%$ pure Bi, Te, and Sb$_{2}$Te$_{3}$ sources. Following substrate outgassing and deposition of a $\sim$50 nm Sb$_{2}$Te$_{3}$ room temperature passivation layer, the substrate temperature was raised to 800 $^{\circ}$C to selectively evaporate the passivation layer, as described elsewhere.\cite{liu_high-quality_2015}  Subsequently, (Bi$_{1-x}$Sb$_{x}$)$_{2}$Te$_{3}$ layers with thicknesses ranging from 5 nm to 34 nm ($\sim$4-31 quintuple layers (QLs)) were grown at 400$^{\circ}$C and annealed in a tellurium flux for > 1 hour. With the Bi:Sb flux ratio controlled by the temperatures of the Bi and Sb effusion cells, we tuned the Bi:Sb fraction to compensate charge carrier conductivity, thereby providing samples that span the crossover regime from n-type to p-type, as summarized in Table I.   

\begin{table}[h!]
\centering
\caption{Thicknesses and compositions of (Bi$_{1-x}$Sb$_{x}$)$_{2}$Te$_{y}$ layers}
\setlength{\tabcolsep}{5pt} 
\setlength{\arrayrulewidth}{0.2mm}  
\begin{tabular}{c c c c c c c c c c c c c}
\noalign{\hrule height 0.3mm} 
\text{Sample} & {Thickness} & \text{Quintuple Layers} & $x_\text{Sb}$ & $y_\text{Te}$  \\
&   \text{(nm)} & \text{(\#)} & $\pm{0.02}$ & $\pm{0.1}$ \\
\hline 
pp      &   5   & 4   & 0.67 & 2.7 \\
pn thin &   20  & 19  & 0.64 & 2.4 \\
pn thick&   34  & 31  & 0.47 & 2.3 \\
nn      &   34  & 31  & 0.61 & 2.6 \\
\noalign{\hrule height 0.3mm} 
\end{tabular}
\label{tab:comp}
\end{table}

\begin{figure*}
\includegraphics[scale=0.75]{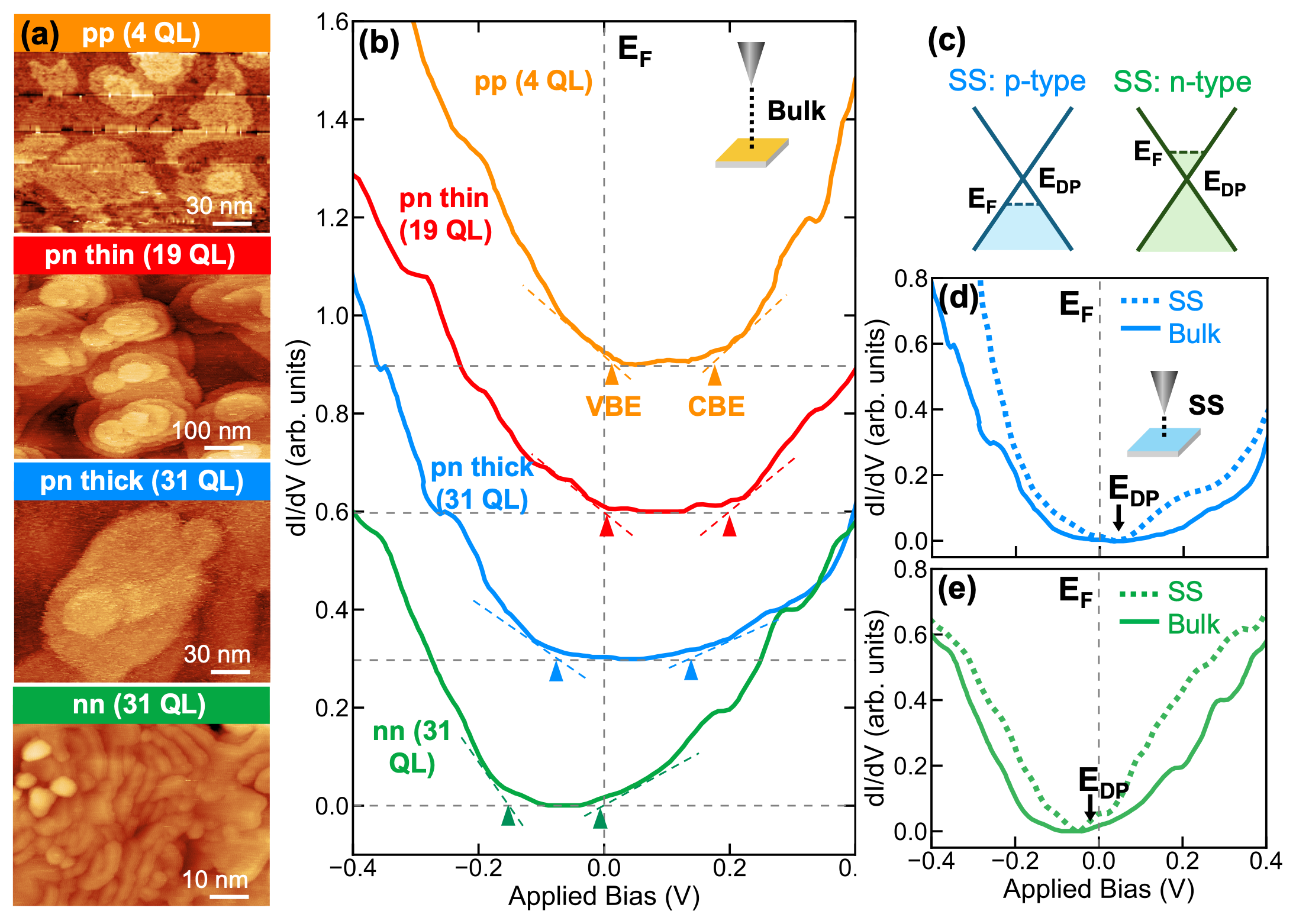}
\caption{\label{fig:epsart} Scanning tunneling microscopy (STM) and scanning tunneling spectroscopy of (Bi$_{1-x}$Sb$_{x}$)$_{2}$Te$_{3}$ films: (a) large-scale STM images of pp ($\Delta z$ = 15.5 nm), pn thin ($\Delta z$ = 9.5 nm), pn thick ($\Delta z$ = 9.8 nm), and nn ($\Delta z$ = 15.8 nm). For all films, large terraces with single quintuple layer (QL) steps are observed, indicating layer-by-layer growth of the van der Waals bonded layers. (b) The differential conductance ($dI/dV$) as a function of bias voltage, corresponding to the energy relative to the Fermi level $E_{F}$. $dI/dV$ for (c) pn thick and (d) nn as a function of bias voltage, corresponding to the energy relative to the $E_{F}$ revealing two distinct states of surface and bulk conduction.}
\label{fig:STM}
\end{figure*}

For all samples, scanning tunneling microscopy (STM) images of the surface topography are shown in Fig.\ \ref{fig:STM}(a). Line-cuts across the images reveal terraces with step heights of $\sim$ 1 nm, as expected for quintuple layer (QL) structures in high-quality layer-by-layer growth of the van der Waals bonded layers. The STM images reveal circular surfaces terraces with sizes ranging from 50 to 100 nm. The number of steps around each terrace was 1-2 (pp) and 3-5 (pn thin, pn thick, and nn). 

To probe the nature of the charge carrier transport and the surface states, we utilize a combination of room temperature scanning tunneling spectroscopy (STS) and variable temperature resistivity and magnetoresistivity (MR) measurements. The STS was performed using a modified variable tip-sample separation method, in which the tip-sample separation is maintained constant during acquisition of a given spectrum, but is adjusted between spectra to enable measurements of bulk and surface states. Because the band gap is significantly larger than $k_B$T (E$_{g}$ $>>$26 meV), when the Dirac point lies within the band gap, the surface states are detected by changing the tip-sample separation that controls tunneling current. STS is more (less) sensitive to states near the Fermi level as the tip is moved closer to (away from) the sample. 

Magnetotransport measurements were performed in a Quantum Design Physical Property Measurement System (PPMS) with the magnetic field applied perpendicular to the surface. Temperatures $<$ 3 K were achieved using a He-3 Refrigerator in tandem with the PPMS. Longitudinal and lateral voltages were measured simultaneously as a function of magnetic field and temperature using standard lock-in techniques. 

Atomistic visualizations were facilitated by the preparation of cross-sectional specimens were fabricated using focused ion beam (FIB) liftout in a Thermo Fisher Scientific Helios G4 Dual Beam system equipped with a Xe plasma ion source. High-angle annular dark-field STEM (HAADF-STEM) and energy dispersive x-ray spectroscopy (EDS) were performed using a Thermo Fisher Spectra 300 Probe-Corrected STEM with a Dual-X EDS detector.

To facilitate further understanding of the twin boundaries, density of states (DOS) calculations were performed using QUANTUM ESPRESSO with electron interactions described by the Projector Augmented Wave (PAW) method.  In addition, the twin boundary conductivity was estimated using a "square lattice" model consisting of twin boundary conducting channels.  Finally, a cubic-lattice version of the Bernevig-Hughes-Zhang tight-binding model was used to compute the twin-boundary modes. More details of the computational methods are described in the supplemental materials. 

\section{Topological Surface States}

For each of the (Bi$_{1-x}$Sb$_{x}$)$_{2}$Te$_{3}$ films, the voltage dependence of the differential conductance, ($dI/dV$), is presented in Fig.\ \ref{fig:STM}(b). For all films, the effective band gaps $E_{g}$ range from 0.15 to 0.22 $\pm$ 0.10 eV, consistent with earlier studies of (Bi$_{1-x}$Sb$_{x}$)$_{2}$Te$_{3}$ alloys.\cite{zhang_band_2011,west_native_2012} For pp and pn thin, the Fermi level ($E_{F}$) is positioned near the valence band edge (VBE), suggesting that p-type carriers are dominant. As the film thickness increases and Sb composition decreases, the $E_{F}$ shifts away from the VBE and towards the conduction band edge (CBE). For pn thick, the $E_{F}$ is closest to the middle of the energy band gap, suggesting compensation between n-type and p-type carriers. For nn,  $E_{F}$ located at the CBE, indicating n-type carriers are dominant.

To probe topological surface states (TSSs) within the films, the tip-sample separation was adjusted to reveal two distinct states of surface (dashed line) and bulk conduction (solid line), as shown in Figs.\ \ref{fig:STM}(c) and (d). It is emphasized that the transport measurements carried out on such samples probe the surface states and not the bulk states. In the energy range of the bulk gap, TSSs exhibit V-shaped spectra, with the zero-conductance point corresponding to the Dirac point ($E_{DP}$), indicated by arrows in Figs.\ \ref{fig:STM}(c) and (d). STS spectra for pn thick shows $E_{DP}$ near 50 meV (Fig.\ \ref{fig:STM}(c)), while for nn, $E_{DP}$ is near -50 meV (Fig.\ \ref{fig:STM}(d)). The shift of the $E_{DP}$ from above to below $E_{F}$ indicates a change in carrier dominance from electrons to holes. Furthermore, the positions of both $E_{DP}$ and $E_{F}$ within the bulk band gap suggest bulk insulating behavior with distinguishable and accessible TSSs.

We now discuss the influence of composition on dominant carrier type in (Bi$_{1-x}$Sb$_{x}$)$_{2}$Te$_{3}$ films. A Sb composition-induced transition from n-type to-p-type conductivity is often reported in (Bi$_{1-x}$Sb$_{x}$)$_{2}$Te$_{3}$ compounds.\cite{kongAmbipolarFieldEffect2011, suSpinChargeConversionManipulated2021} Similarly, an n-to-p transition with increasing Sb composition is apparent for our nn, pn thin, and pp films (Fig.\ \ref{fig:STM}(b)). However, for pn thick, in spite of its lower Sb composition, the Fermi level has shifted towards midgap, presumably due to hole generation induced by antisite defects such as Bi$_{Te}$ and Sb$_{Te}$ in (Bi$_{1-x}$Sb$_{x}$)$_{2}$Te$_{3}$\cite{zhaoExcellentThermoelectricPerformance2021, lostak_antisite_1994} resulting in carrier compensation.

\begin{figure}
    \includegraphics[width=\linewidth]{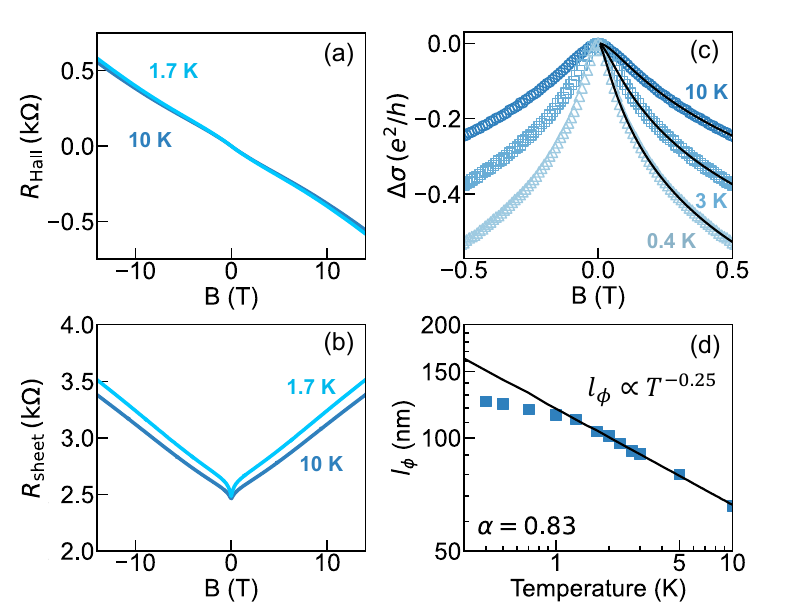}
    \caption{Magnetotransport data for nn (31 QL): (a) Hall resistance and (b) sheet resistance as a function of magnetic field at 10 K and 1.7 K. (c) Magnetoconductivity as a function of temperature (open symbols) with Hikami-Larkin-Nagaoka fit (solid lines). (d) Phase coherence length as a function of temperature (square symbols) and a fit proportional to $T^{-0.25}$ (solid line). At $T = 0.4$ K, we calculate $\alpha = 0.83$.}
    \label{fig:transport}
\end{figure}

To explore conduction through TSSs, we consider magnetotransport data in the cryogenic regime.  Figs.\ \ref{fig:transport}(a) and (b) show example Hall and sheet resistances for pn thick (31 QL), as a function of magnetic field, for a variety of temperatures. Due to the negative Hall coefficient in Fig.\ \ref{fig:transport}(b), the majority carriers are electrons. The low-field Hall data and the zero-field sheet resistance were used to compute surface state carrier concentrations that are consistent with literature reports,\cite{liu_high-quality_2015, weyrichGrowthCharacterizationTransport2016} as summarized in Table \ref{tab:transport summary}.

For temperatures, $T$, below 10 K, the Hall resistance (Fig.\ \ref{fig:transport}(a)) is essentially temperature-independent and decreases monotonically with magnetic field, $B$. Meanwhile, the sheet resistance (Fig.\ \ref{fig:transport}(b)) exhibits a cusp-like decrease near $B = 0$ that becomes more pronounced at the lowest temperatures. This cusp-like behavior suggests the presence of weak anti-localization (WAL).\cite{bergmann_weak_1982} Therefore, using a matrix inversion of the Hall and sheet resistances, we consider the B-field dependence of normalized longitudinal magnetoconductivity $\Delta \sigma = \sigma(B) - \sigma(B=0)$, shown in Fig.\ \ref{fig:transport}(c).  Quantum corrections to the conductivity are estimated by the Hikami-Larkin-Nagaoka formula:\cite{hikami_spin-orbit_1980}
\begin{equation}
    \Delta \sigma = -\frac{\alpha e^2}{\pi h} \bigg[\Psi\bigg(\frac{1}{2} + \frac{B_\phi}{B}\bigg) - \ln \bigg(\frac{B_\phi}{B}\bigg)\bigg]
    \label{eqn: HLN}
\end{equation}
where $\alpha$ is the WAL factor (an estimate of the number of conducting channels), $e$ is the electron charge, $h$ is the Planck constant, $\Psi$ is the digamma function, and $B_\phi = h/(8\pi e l_\phi^2)$ is the dephasing field, where $l_\phi$ is the phase coherence length, i.e. the distance carriers travel before losing phase information due to inelastic scattering.

\begin{figure}[h]
    \centering
    \includegraphics[width=0.8\linewidth]{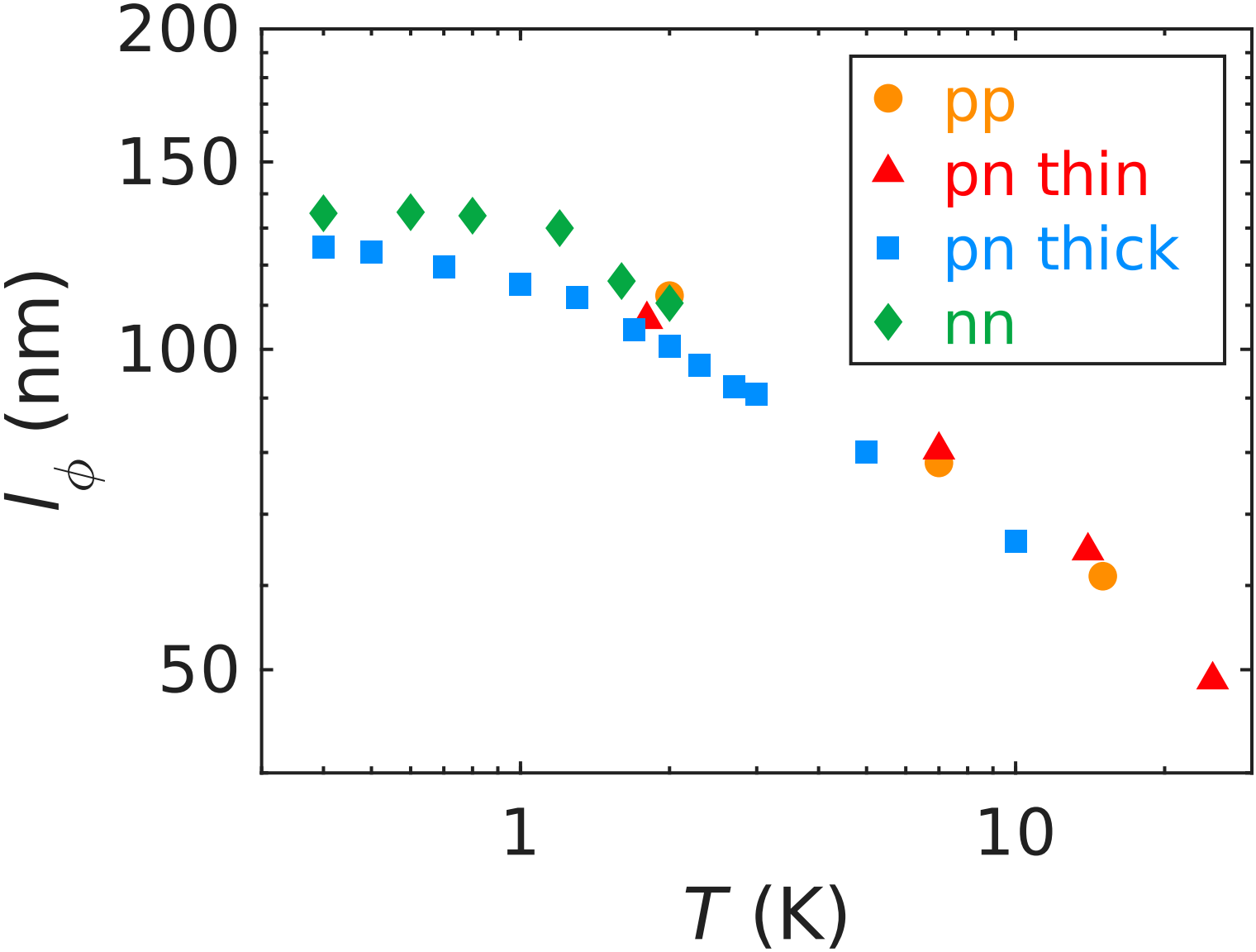}
    \caption{Phase coherence lengths, $l_\phi$, as a function of temperature for the pp (4 QLs), pn thin (19 QLs), pn thick (31 QLs), and nn (31 QLs). For $T > 1$ K, similar $T$-dependencies are apparent for all samples.}
    \label{fig: supp lphi all}
\end{figure}

For $T> 1.5$ K, fits to $\Delta \sigma$ were performed using the $\alpha$ value computed at low temperature, with $B_\phi$ as the only fit parameter. For the pn thick sample, the resulting $T$-dependence of $l_\phi$ is shown in Fig.\ \ref{fig:transport}(d). Meanwhile, for all samples, the corresponding temperature dependences of $l_\phi$ are presented in Fig.\ \ref{fig: supp lphi all}, with the values of $\alpha$, $l_\phi$, and $p$ summarized in Table \ref{tab:transport summary}. For temperatures above 1 K, all samples exhibit similar temperature dependence of the phase coherence lengths.

\begin{table} [b]
\centering
\caption{Low-field and high-field transport parameters for (Bi$_{1-x}$Sb$_{x}$)$_{2}$Te$_{3}$ films at low temperatures, $T$, including the carrier concentration, $n$, the carrier mobility, $\mu$, the weak anti-localization factor, $\alpha$, the phase coherence length, $l_\phi$, and the parameter $p$. Values for pp (4 QLs) were obtained from Ref.~\onlinecite{liu_origins_2016}}.
\setlength{\tabcolsep}{3pt} 
\setlength{\arrayrulewidth}{0.2mm}  
\begin{tabular}{c c c c c c c}
    \noalign{\hrule height 0.3mm} 
    \makecell{\text{Sample} \\} & \makecell{$T$ \\ (K)} & \makecell{ $n_{2D}$ \\ ($10^{13}$ cm$^{-2}$) } & \makecell{$\mu$ \\ (cm$^2$/V$\cdot$s)} & \makecell{ $\alpha$ \\  } & \makecell{ $l_\phi$ \\ (nm) } &  \makecell{ $p$ \\  } \\
    \hline 
    pp       & 2   & 0.5  & 300  & 1    & 110 & 0.60 \\
    pn thin  & 1.8 & 0.5  & 270  & 1.77 & 106 & 0.48 \\
    pn thick  & 0.4 & 1.5  & 166 & 0.83 & 125 & 0.50 \\
    nn        & 0.4 & 1.1  & 170 & 0.78 & 135 & 0.42 \\
    \noalign{\hrule height 0.3mm} 
\end{tabular}
\label{tab:transport summary}
\end{table}

Interestingly, $l_\phi$ increases monotonically with decreasing temperature. Fitting of $l_\phi(T)$ to a power law of the form $T^{-p/2}$ yields $p=0.5$, similar to the $p$ value attributed to coupling between TSSs and so-called localized charge puddles associated with defects in thin films.\cite{liao_enhanced_2017,beidenkopf_spatial_2011} Furthermore, it has been reported that $\alpha = 0.5$ corresponds to a single conduction channel,\cite{chaWeakAntilocalizationBi2SexTe1x32012,urkudeTemperatureImpurityEffect2017} while $\alpha > 0.5$ indicates contributions from top and bottom surfaces or decoupling between surface states and the bulk.\cite{steinbergElectricallyTunableSurfacebulk2011,leThicknessdependentMagnetotransportProperties2017,yuRobustGaplessSurface2020} The variation of $\alpha$ observed in (Bi$_{1-x}$Sb$_{x}$)$_{2}$Te$_{3}$ films (Table \ref{tab:transport summary}) suggests the presence of unidentified charge sources beyond the top surface states may be influencing conduction channels.

\section{Atomic Structure and Chemistry}

\begin{figure*}
    \centering
    \includegraphics[width=0.75\textwidth]{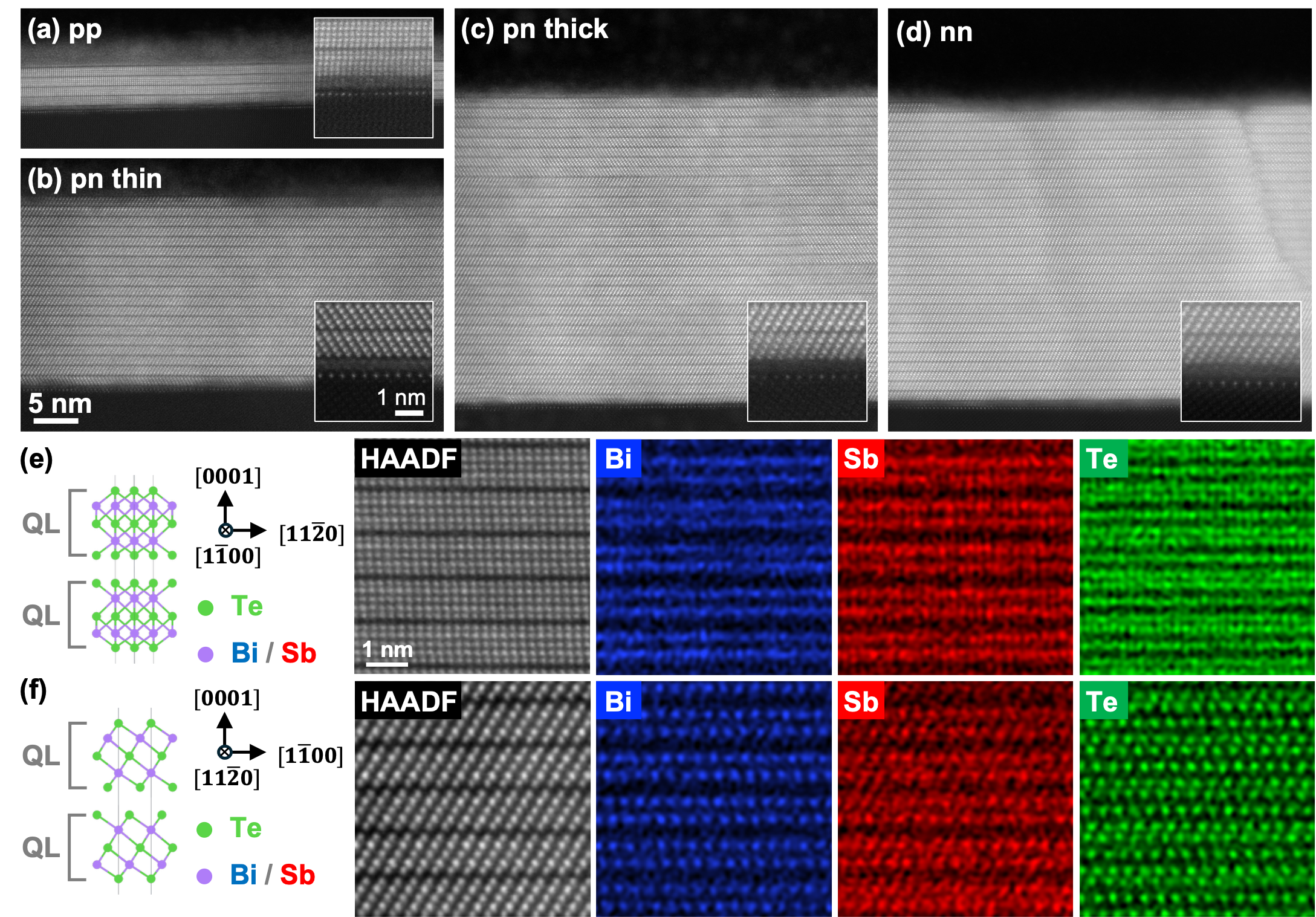} 
    \caption{High-angle annular dark-field scanning transmission electron micrographs (HAADF-STEM) revealing quintuple layers (QLs) within (Bi$_{1-x}$Sb$_{x}$)$_{2}$Te$_{3}$ films: (a) $[1\overline{1}00]$ projection of pp (4 QL); $[11\overline{2}0]$ projections of (b) pn thin (19 QL), (c) pn thick (31 QL) and (d) nn (31 QL). The QLs consist of 5 atomic layers, separated by van der Waals gaps. High resolution (e) $[1\overline{1}00]$ and (f) $[11\overline{2}0]$ projections of the atomic structures, HAADF-STEM images, and energy-dispersive x-ray spectroscopy elemental maps of Bi (blue), Sb (red), and Te (green). }
    \label{fig:TEM_thickness}
\end{figure*}

We now consider the atomic structures of the (Bi$_{1-x}$Sb$_{x}$)$_{2}$Te$_{3}$ layers and their interfaces with the sapphire substrate. Figure \ref{fig:TEM_thickness} shows high-angle annular dark-field scanning transmission electron micrographs (HAADF-STEM) of the (a) pp (4 QLs), (b) pn thin (19 QLs), (c) pn thick (31 QLs), and (d) nn (31 QLs) samples, with high magnification views of the (Bi$_{1-x}$Sb$_{x}$)$_{2}$Te$_{3}$/Al$_{2}$O$_{3}$ interfaces shown as insets. Each QL consists of 5 atomic layers, with multiple QLs separated by van der Waals gaps. Interestingly, horizontal red arrows point to examples of ordered atomic layers apparent on the substrate side of the (Bi$_{1-x}$Sb$_{x}$)$_{2}$Te$_{3}$/Al$_{2}$O$_{3}$ interface, as will be discussed in more detail below.

Example HAADF images and corresponding atomic EDS mappings for $[1\overline{1}00]$ and $[11\overline{2}0]$ projections of multiple QLs of (Bi$_{1-x}$Sb$_{x}$)$_{2}$Te$_{3}$ are shown in Figs.\ \ref{fig:TEM_thickness}(e) and \ref{fig:TEM_thickness}(f). For the $[1\overline{1}00]$ projection in Fig.\ \ref{fig:TEM_thickness}(e), vertical alignment of the five atomic layers within each QL is observed.  On the other hand, for the $[11\overline{2}0]$ projection in Fig.\ \ref{fig:TEM_thickness}(f), ~30$^{\circ}$ misalignment of the five atomic layers within each QL is apparent. Notably, in both projections, the blue (Bi) and red (Sb) dots in the energy dispersive spectroscopy (EDS) mappings align with the second and the fourth bright dots in each QL of the HAADF images, whereas the green (Te) dots correspond to the first, third, and fifth bright dots in each QL. Schematic diagrams of the atomic layer structures are shown in the left panels of Figs.\ \ref{fig:TEM_thickness}(e) and (f), illustrating the alternating sequence of Te and Bi/Sb within each QL. As summarized in Table \ref{tab:comp}, the (BiSb):Te ratios range from 2:2.3 to 2:2.7. The slight deviations from the ideal (BiSb):Te ratio of 2:3 is likely due to the presence of Bi on Te site (Bi$_{Te}$), Sb on Te site (Sb$_{Te}$), or Te vacancies (V$_{Te}$).

\begin{figure}[b]
    \centering
    \includegraphics[width=0.5\textwidth]{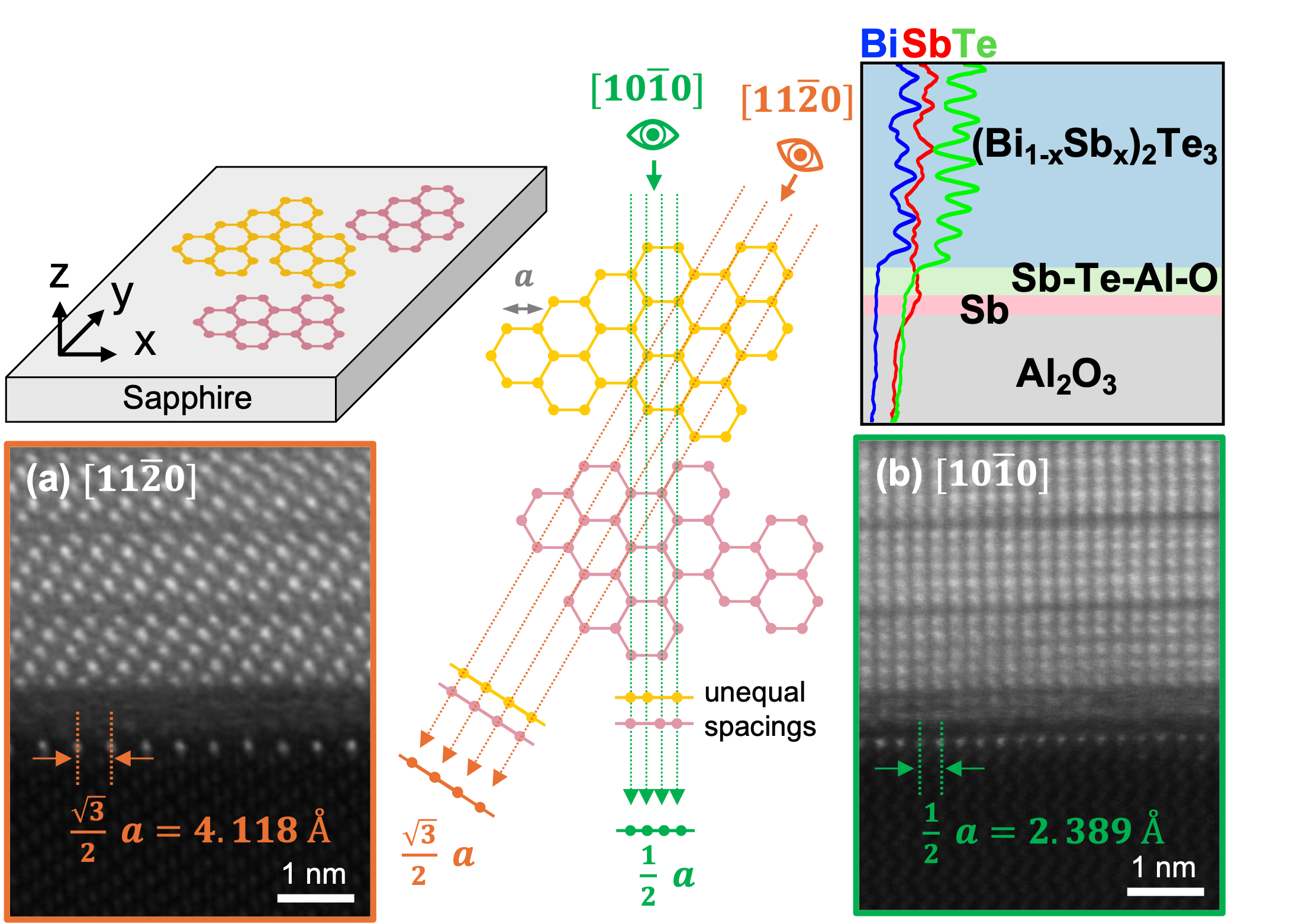} 
    \caption{Example high-angle annular dark-field scanning transmission electron micrographs (HAADF-STEM) in the vicinity of the (Bi$_{1-x}$Sb$_{x}$)$_{2}$Te$_{3}$/Al$_{2}$O$_{3}$ interface. For the (a) $[11\overline{2}0]$  and (b) $[10\overline{1}0]$ projections, constant Sb-Sb separations are 4.118 Å and 2.389 Å, respectively. For a single domain antimonene monolayer, illustrated in yellow, two distinct Sb-Sb separations are apparent in the $[10\overline{1}0]$ projection. Since a constant Sb-Sb separation is instead observed, multiple domains of antimonene monolayers, illustrated in yellow and pink, are expected.}
    \label{fig: Sb monolayer}
\end{figure}

HAADF imaging and EDS maps in the vicinity of the (Bi$_{1-x}$Sb$_{x}$)$_{2}$Te$_{3}$/Al$_{2}$O$_{3}$ interfaces (Fig.\ \ref{fig: Sb monolayer}) reveal a crystallized atomic layer of Sb and a disordered interfacial layer containing Al, O, Sb, and Te. These interlayers are likely associated with the passivation layer that facilitates high quality growth of tetradymite films.\cite{liu_high-quality_2015} The crystallized atomic layers of Sb observed in all of the films primarily consist of antimonene monolayers, with antimonene bi-layers observed in $<$ 10\% of the imaged regions. 

We now consider the structure of the antimonene monolayers based upon HAADF-STEM imaging in the $[11\overline{2}0]$ and $[10\overline{1}0]$ projections. In Figs.\ \ref{fig: Sb monolayer}(a) and \ref{fig: Sb monolayer}(b), the separation between Sb atoms in the $[11\overline{2}0]$ and $[10\overline{1}0]$ projections is 4.118 Å and 2.389 Å, respectively. Due to the honeycomb structure of the antimonene monolayer,\cite{shaoEpitaxialGrowthFlat2018} Sb-Sb atomic separations of \(\frac{\sqrt{3}}{2}\)a, where a is the Sb-Sb nearest-neighbor separation, are expected in the $[11\overline{2}0]$ projection.  Meanwhile, for the $[10\overline{1}0]$ projection, two distinct Sb-Sb atomic separations, a and \(\frac{1}{2}\)a, are expected. Since our antimonene monolayers appear to contain fixed values of Sb-Sb atomic separations in both the $[11\overline{2}0]$ and $[10\overline{1}0]$ projections, we consider the possibility that two or more domains have formed, depicted as pink and yellow honeycomb structures in the middle of Fig.\ \ref{fig: Sb monolayer}.  For a monolayer with both domains, the average Sb-Sb atomic separation is 4.118 $\text{Å} \, (\approx \frac{\sqrt{3}}{2}a)$ in $[11\overline{2}0]$ direction and 2.389 $\text{Å} \, (\approx \frac{1}{2}a)$ in $[10\overline{1}0]$ direction. 

\section{Twin Boundaries}

As shown in Figs.\ \ref{fig: DOS}(a)-(e), extended defects, including basal twins and $60^\circ$ twins, are observed in the $[11\overline{2}0]$ projections of (Bi$_{1-x}$Sb$_{x}$)$_{2}$Te$_{3}$ thin films. At a basal twin boundary (Fig.\ \ref{fig: DOS}(a)), two adjacent QLs are rotated by $180^\circ$ along the c-axis, resulting in a mirror plane (orange dashed line) parallel to the surface. On the other hand, at a $60^\circ$ twin boundary (blue dashed line in Figs.\ \ref{fig: DOS}(b)-(e)), the mirror plane is perpendicular to the surface, and the atomic structures within the QLs are mirrored across the basal plane. Both basal twins and $60^\circ$ twins are observed at various heights, with densities of $\sim$$10^{10}\ \text{cm}^{-2}$ for all (Bi$_{1-x}$Sb$_{x}$)$_{2}$Te$_{3}$ films.


The corresponding computed DOS for (Bi$_{1-x}$Sb$_{x}$)$_{2}$Te$_{3}$ in the bulk, and in the vicinity of the basal and $60^\circ$ twins, are shown in Fig.\ \ref{fig: DOS}. For both the bulk and in the vicinity of basal twins (Figs.\ \ref{fig: DOS}(f) and (g)), the DOS exhibits negligible weight near the Fermi energy (black vertical dotted line). In contrast, in Fig.\ \ref{fig: DOS}(h), the $60^\circ$ twin shows a significant DOS in the bandgap, suggesting that the $60^\circ$ twin boundaries contribute a 2D carrier density of $3.22\times10^{13}\ \rm{cm}^{-2}$, consistent with earlier reports on Bi$_{2}$Te$_{3}$.\cite{kimFreeelectronCreation60deg2016} Thus, $60^\circ$ twins in (Bi$_{1-x}$Sb$_{x}$)$_{2}$Te$_{3}$ are expected to play a key role in carrier transport.

\begin{figure*}
    \includegraphics[scale=0.65]{Figures/BiSbTe-twins-both.png}
    \caption{High-angle annular dark-field scanning transmission electron micrographs (HAADF-STEM) of (a) basal twin, (b) $60^\circ$ twin at the interface, (c) $60^\circ$ twin at the surface, (d) $60^\circ$ twin in the middle of the film, and (e) $60^\circ$ twin across all quintuple layers (QLs), respectively. Two adjacent QLs are rotated by $180^\circ$ along the c-axis at the basal twin boundary, while the atomic structures within the QLs are mirrored across the basal plane at the $60^\circ$ twin boundary. The orange and blue dashed lines mark the boundaries of the basal and $60^\circ$ twin, respectively. Density of states (DOS) per unit volume of BiSbTe$_3$ for (f) bulk, (g) basal twin, and (h) $60^\circ$ twin structures. The insets present the atomic structures used in the DFT calculations. In (g), the  orange dashed line represents the basal twin boundary, while in (h), the blue dashed line indicates the $60^\circ$ twin boundary. The black vertical dotted lines indicate the position of the Fermi energy, $E_{F}$.}
    \label{fig: DOS}
\end{figure*}

We now consider conduction through the twin boundaries. We define the intrinsic conductivity of a $60^\circ$ twin boundary as $\sigma_{twin} =  n_{twin} e \mu_{twin}$, where $n_{twin}$ is the two-dimensional carrier concentration and $\mu_{twin}$is the carrier mobility, both associated with the boundary. We note that this analysis is independent of the sign of the carriers. 

\begin{figure}
    \centering
    \includegraphics[width=0.95\linewidth]{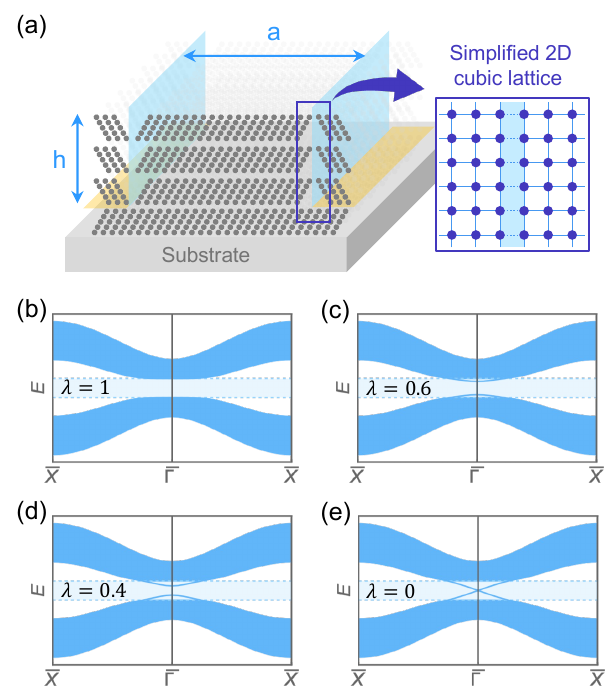}
    \caption{(a) Left panel: Schematic of twin boundary model (not drawn to scale) of BiSbTe$_3$. Basal twins (orange) and $60^\circ$ twins (blue) are parallel and perpendicular to the substrate surface, respectively. $60^\circ$ twins are separated by a distance $a$ and have a depth $h$. Right panel: Illustration of a simplified twin boundary (light blue region) between two domains. (b-e) Band structure of the twin boundary modes, displaying the dispersion along a 1D path $\bar{X}$-$\bar{\Gamma}$-$\bar{X}$, where $\bar{\Gamma}$ and $\bar{X}$ correspond to the momentum points $(0,0)$ and $(\pi,0)$ of the twin-boundary Brillouin zone, respectively. The bulk energy spectra are represented by thick blue parabolic-like bands, with light blue regions between the dashed blue horizontal lines corresponding to the bandgaps, and narrow blue lines corresponding to emerging in-gap states localized at the twin boundaries. The twin boundary tunneling probability is quantified by a factor $\lambda$ that ranges from 0 to 1, with $\lambda=0$ corresponding to full suppression of tunneling through the twin boundary. In the limit of fully suppressed twin boundary tunneling, the band dispersion transforms into a massless Dirac cone with topologically protected states localized at the twin boundaries, as illustrated in (e).}
    \label{fig: twin bndy model}
\end{figure}

To compute the twin boundary conductivity, we utilize a ``square lattice'' model consisting of an $L \times L$ grid of perpendicularly intersecting 60$^\circ$ twin boundaries, with spacing between twins, $a$, and depth of twins, $h$, as shown in Fig.\ \ref{fig: twin bndy model}(a). The twin boundaries are treated as conducting ``channels'' with a width $w$ that is taken to zero in the final calculation. For $N$ twin boundaries, with $N = L/a$, the resistance of each twin is $R = \rho L/(hw)$.  Dividing by the number of twins $N$, the sheet resistance, $R_s$, can be expressed as
\begin{equation}
    R_s = \rho \frac{a}{hw} = \frac{1}{ne\mu} \frac{a}{hw},
    \label{eqn: sqmod sheet res}
\end{equation}
where the twin boundary resistivity $\rho$ is replaced by $1/(ne\mu$). Here, $n$ is a three-dimensional carrier concentration of the twin boundary channel and is related to the two-dimensional carrier concentration of the twin via $n_{twin} = nw$.

For the Hall resistance, we consider a twin boundary with a current $i$ flowing through it in the $\hat{x}$ direction. A perpendicular magnetic field $\mathbf{B} = B\hat{z}$ produces a Hall voltage $V_{H}^{twin} = \mathcal{E}_{\perp} w$ across the boundary where $\mathcal{E}_{\perp}$ is the Hall field. For an $N \times N$ grid of twin boundaries, the total Hall voltage becomes:
\begin{equation}
    V_{H} = V_{H}^{twin} N = \mathcal{E}_{\perp} w N.
    \label{eqn: sqmod hall 1}
\end{equation}

For $N$ twin boundaries, the total current is $I = iN$. The current density is then expressed in terms of the drift velocity, $v_d$, and the 2D carrier concentration, $n_{twin}$ of the twin, as follows:
\begin{equation}
    J = \frac{I}{\text{total area}} = \frac{I}{hwN} = n_{twin}ev_d,
    \label{eqn: current density}
\end{equation}
Since the current flows in the $\hat{x}$ direction, the Lorentz force must be zero at twin boundary crossings:
\begin{equation}
    ev_d B = e\mathcal{E}_{\perp}.
    \label{eqn: lorentz zero}
\end{equation}
Combining Eqs.\ \eqref{eqn: sqmod hall 1}, \eqref{eqn: current density}, and \eqref{eqn: lorentz zero} yields the Hall resistance:
\begin{equation}
    R_{Hall} =\frac{B}{n_{twin}eh}.
    \label{eqn: sqmod Hall res}
\end{equation}

The diagonal and off-diagonal components of twin conductivity can be calculated via matrix inversion using Eqs.\ \eqref{eqn: sqmod sheet res} and \eqref{eqn: sqmod Hall res}:
\begin{subequations}
    \begin{align}
        \sigma_{xx}^{twin}(B) &= \frac{n_{twin}e \mu (h/a)}{1 + [\mu B (w/a)]^2}, \\
        \sigma_{xy}^{twin}(B) &= -\frac{n_{twin}e \mu^2B (h/a) (w/a)}{1 + [\mu B (w/a)]^2}.
    \end{align}
\end{subequations}
In the limit where $w \ll a$ (i.e. $w$ is taken to 0), the equations above simplify to:
\begin{subequations}
    \begin{align}
        \sigma_{xx}^{twin}(B) &= \sigma_{xx}^{twin} (h/a), \\
        \sigma_{xy}^{twin}(B) &= 0.
    \end{align}
\end{subequations}
where $\sigma_{xx}^{twin} = n_{twin}e\mu_{twin}$ is the twin conductivity. 

In our ``square lattice'' model, $60^\circ$ twin boundaries only possess diagonal components of conductivity. We can therefore drop the subscript on twin conductivity and define $\sigma_{xx}^{twin} = \sigma_{twin}$. 
Furthermore, the magnetic-field induced Lorentz force appears solely at the intersections between twin boundaries. Thus, any carriers localized at the $60^\circ$ twin boundaries contribute to the longitudinal conductivity, as follows:
\begin{subequations}
    \label{eqn: conductivity + twin}
    \begin{align}
        \sigma_{xx}(B) &= \frac{n e \mu}{1 + \mu^2B^2} + \sigma_{twin}\bigg(\frac{h}{a}\bigg) \label{eqn: sigmaxx} \\
        \sigma_{xy}(B) & = -\frac{ne\mu^2B}{1 + \mu^2B^2}, \label{eqn: sigmaxy}
    \end{align}
\end{subequations}
where $n$ is the two-dimensional carrier concentration and $\mu$ is the effective carrier mobility associated with surface states unrelated to the boundary. 

To investigate the effects of $\sigma_{twin}(h/a)$ on carrier transport, we can use Eqs.\ \eqref{eqn: sigmaxx} and \eqref{eqn: sigmaxy} to solve for mobility $\mu$, which can be expressed as:
\begin{equation}
    \mu = -\frac{\sigma_{xy}(B)}{B\bigg[\sigma_{xx}(B) - \sigma_{twin}\bigg(\dfrac{h}{a}\bigg)\bigg]}.
    \label{eqn: supp mobility}
\end{equation}

To determine the geometric factor $(h/a)$, we approximate the twin boundary separation, $a$, as the square root of the inverse of the twin boundary density, i.e. $a \sim$ 100 nm. We then utilize the HAADF images from Fig.\ \ref{fig: DOS} to determine $h$ ($\sim$ 10 nm), leading to $(h/a) \sim 0.1$. 

\begin{figure}
    \centering
    \includegraphics[width=0.85\linewidth]{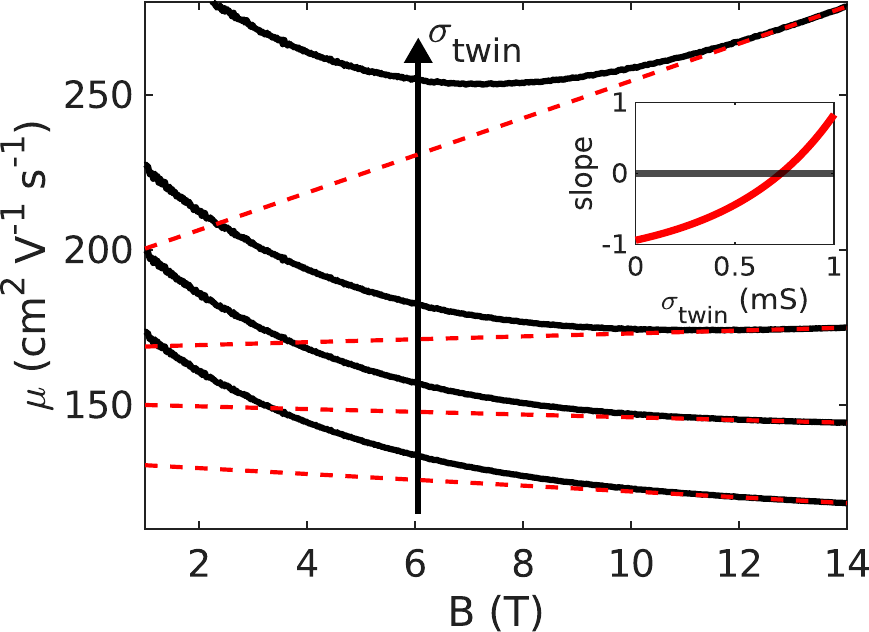}
    \caption{Example calculation of the upper bound for $\sigma_{twin}$ for pn thick (31 QL). The mobility, $\mu$, is shown as a function of magnetic field for various $\sigma_{twin}$ (solid lines), and the slope of $\mu$ between 13 T and 14 T is shown in dashed red lines. In the inset, we plot the value of the slope (arbitrary units) as a function of $\sigma_{twin}$. We see that the slope becomes positive (i.e. $\mu$ begins to increase) when $\sigma_{twin} > 0.73$ mS.}
    \label{fig: supp up bound}
\end{figure}

Figure \ref{fig: supp up bound} shows $\mu(B)$ for various $\sigma_{twin}$ values, with $(h/a) = 0.1$. Typically, $\mu$ is expected to decrease monotonically with magnetic field.\cite{hattori_note_1960,matz_non-ohmic_1964} However, for $\sigma_{twin} > 0.73$ mS, we see that $\mu$ begins to increase with magnetic field, which is unphysical. This value for $\sigma_{twin}$ is found by analyzing the slope of $\mu$ from 13 T to 14 T (see inset of Fig.\ \ref{fig: supp up bound}). Therefore, $\sigma_{twin} \leq 7.3 \times 10^{-4}$ S, which yields yields a carrier mobility as high as $\mu_{twin} \sim 142$ cm$^2$/(V$\cdot$s).

To compute the in-gap states induced by the twin boundaries, we consider their impact on electron tunneling between the two domains that they separate. We use a cubic-lattice version of the Bernevig-Hughes-Zhang model,\cite{bernevig_BHZ_model2006} as illustrated in Fig.\ \ref{fig: twin bndy model}(a). The twin boundary tunneling probability is quantified by a factor $\lambda$ that ranges from 0 to 1, with $\lambda=0$ corresponding to full suppression of tunneling through the twin boundary. If the appearance of a domain boundary increases the distance between atoms (on the two sides of the domain wall), then it suppresses the tunneling ($\lambda=0$). The larger than separation, the stronger the suppression of tunneling (smaller lambda). The increasing of the separation leads to decreasing $\lambda$. For the twin boundary in the 2D unit cell (Figs.\ \ref{fig: twin bndy model}(b)-(e)), the energy dispersion is computed along a 1D path $\bar{X}$-$\bar{\Gamma}$-$\bar{X}$, where $\bar{\Gamma}$ and $\bar{X}$ correspond to the momentum points $(0,0)$ and $(\pi,0)$ of the twin-boundary Brillouin zone, respectively.

For $\lambda=1$, the twin boundary tunneling probability is highest, resulting in an energy gap in the bulk dispersion. However, as twin boundary tunneling is suppressed $\lambda=0.6$ and 0.4 the mass (i.e., energy gap) of the Dirac cone decreases, with the emergence of in-gap states, as illustrated by indicated by narrow blue lines in Figs.\ \ref{fig: twin bndy model}(c) and (d). Finally, for $\lambda=0$, the twin boundary tunneling probability is negligible, and the twin-boundary states evolve into topologically protected surface states, with their dispersion forming a gapless Dirac cone (Fig.\ \ref{fig: twin bndy model}(e)). In this case, $60^\circ$ twin boundaries would host TSSs that contribute to the total conductivity.

\section{Summary and Outlook}

In summary, we have investigated conduction paths associated with extended defects in (Bi$_{1-x}$Sb$_{x}$)$_{2}$Te$_{3}$ thin films. Our calculations suggest an enhanced DOS near the Fermi level within $60^\circ$ twin boundaries with computed 2D carrier density in excess of $3\times10^{13}\ \rm{cm}^{-2}$. Additionally, we presented an isotropic model for twin boundaries which allowed us to obtain an expression for twin boundary conductivity $\sigma_{twin}$ and place an upper bound of approximately $7.3\times10^{-4}$ S. Combined with the computed 2D carrier density generated by $60^\circ$ twin boundaries, the estimated $\mu_{twin} \sim 142$ cm$^2$/(V$\cdot$s). Finally, we discuss the role of twin boundaries in facilitating a transition from a massive Dirac cone dispersion to gapless, topologically protected surface states. These results provide a new understanding of the contributions of twin boundaries to carrier conduction in non-degenerate TIs, including the possibility of contributing topological states.


\section*{Supplementary Material}

In the supplementary material, we present experimental and computational methods including Hall bar contact misalignment corrections, analysis of STS data, DOS calculations, and tight-binding model calculations of twin boundary modes.


\section*{Acknowledgments}
This research was supported by the Air Force Office of Scientific Research through the Multidisciplinary University Research Initiative, Award No. FA9550-23-1-0334. The authors acknowledge the assistance of the staff at the Michigan Center for Materials Characterization at the University of Michigan.

\section*{Author Declarations}

\subsection*{Conflict of Interest}
The authors have no conflicts to disclose.

\subsection*{Author Contributions}
\textbf{Abby Liu}: Data Curation (equal); Investigation (equal); Visualization (lead); Writing (equal); Review and editing (equal)
\textbf{Armando Gil}: Data Curation (equal); Investigation (equal); Writing (supporting);
\textbf{Moon-ki Choi}: Formal analysis (supporting); Methodology (equal); Visualization (supporting); Writing (supporting);
\textbf{Berna Akgenc Hanedar}: Methodology (equal); Visualization (supporting); Writing (supporting);
\textbf{Zecheng You}: Formal analysis (supporting); Writing (supporting);
\textbf{Shriya Sinha}: Investigation (supporting); Writing (supporting);
\textbf{Tahsin Hakioglu}: Methodology (equal); Review and editing (equal); Writing (supporting);
\textbf{Harley T. Johnson}: Concept (equal); Funding acquistion (equal); Supervision (equal); Writing (equal); Review and editing (equal) 
\textbf{Kai Sun}: Concept (equal); Funding acquistion (equal); Supervision (equal); Writing (equal); Review and editing (equal)
\textbf{Roy Clarke}: Concept (equal); Funding acquistion (equal); Supervision (equal); Writing (equal); Review and editing (equal)
\textbf{Ctirad Uher}: Concept (equal); Funding acquistion (equal); Supervision (equal); Writing (equal); Review and editing (equal)
\textbf{Cagliyan Kurdak}: Concept (equal); Funding acquistion (equal); Supervision (equal); Writing (equal); Review and editing (equal)
\textbf{Rachel S. Goldman}: Concept (equal); Funding acquistion (equal); Project Administration (lead); Supervision (equal); Writing (equal); Review and editing (lead) 

\section*{Data Availability}
The data that support the findings of this study are available within the article and its supplementary material.


\section*{References}
\bibliography{References} 

@article{bernevig_BHZ_model2006,
title = {Quantum Spin {Hall} Effect and Topological Phase Transition in {HgTe} Quantum Wells},
	volume = {314},
	url = {https://www.science.org/doi/10.1126/science.1133734},
	doi = {DOI: 10.1126/science.1133734},
	journal = {Sci.},
	author = {Bernevig, B. Andrei and Hughes, Taylor L. and Zhang, Shou-Cheng},
	month = Dec,
	year = {2006},
	pages = {1757},
}

@article{maciejko_quantum_2011,
	title = {The Quantum Spin {Hall} Effect},
	volume = {2},
	issn = {1947-5454, 1947-5462},
	url = {https://www.annualreviews.org/doi/10.1146/annurev-conmatphys-062910-140538},
	doi = {10.1146/annurev-conmatphys-062910-140538},
	number = {1},
	urldate = {2024-04-11},
	journal = {Annu. Rev. Condens. Matter Phys.},
	author = {Maciejko, Joseph and Hughes, Taylor L. and Zhang, Shou-Cheng},
	month = mar,
	year = {2011},
	pages = {31},
}

@article{zhang_experimental_2009,
	title = {Experimental Demonstration of Topological Surface States Protected by Time-Reversal Symmetry},
	volume = {103},
	copyright = {http://link.aps.org/licenses/aps-default-license},
	issn = {0031-9007, 1079-7114},
	url = {https://link.aps.org/doi/10.1103/PhysRevLett.103.266803},
	doi = {10.1103/PhysRevLett.103.266803},
	number = {26},
	urldate = {2024-04-11},
	journal = {Phys. Rev. Lett.},
	author = {Zhang, Tong and Cheng, Peng and Chen, Xi and Jia, Jin-Feng and Ma, Xucun and He, Ke and Wang, Lili and Zhang, Haijun and Dai, Xi and Fang, Zhong and Xie, Xincheng and Xue, Qi-Kun},
	month = dec,
	year = {2009},
	pages = {266803},
}

@article{west_native_2012,
	title = {Native defects in second-generation topological insulators: {Effect} of spin-orbit interaction on {Bi$_{2}$Se$_{3}$}},
	volume = {86},
	copyright = {http://link.aps.org/licenses/aps-default-license},
	issn = {1098-0121, 1550-235X},
	shorttitle = {Native defects in second-generation topological insulators},
	url = {https://link.aps.org/doi/10.1103/PhysRevB.86.121201},
	doi = {10.1103/PhysRevB.86.121201},
	number = {12},
	urldate = {2024-04-11},
	journal = {Phys. Rev. B},
	author = {West, D. and Sun, Y. Y. and Wang, Han and Bang, Junhyeok and Zhang, S. B.},
	month = sep,
	year = {2012},
	pages = {121201},
}

@article{lostak_antisite_1994,
	title = {Antisite defects in {BiSbTe$_{3}$} crystals doped with indium atoms},
	volume = {143},
	issn = {1521-396X},
	url = {https://onlinelibrary.wiley.com/doi/abs/10.1002/pssa.2211430210},
	doi = {10.1002/pssa.2211430210},
	number = {2},
	urldate = {2024-04-11},
	journal = {Phys. Status Solidi A},
	author = {Lošťák, P. and Karamazov, S. and Horák, J.},
	year = {1994},
	pages = {271},
}

@article{hsieh_observation_2009,
	title = {Observation of {Time}-{Reversal}-{Protected} {Single}-{Dirac}-{Cone} {Topological}-{Insulator} {States} in {Bi$_{2}$Te$_{3}$} and {Sb$_{2}$Te$_{3}$}},
	volume = {103},
	copyright = {http://link.aps.org/licenses/aps-default-license},
	issn = {0031-9007, 1079-7114},
	url = {https://link.aps.org/doi/10.1103/PhysRevLett.103.146401},
	doi = {10.1103/PhysRevLett.103.146401},
	number = {14},
	urldate = {2024-04-11},
	journal = {Phys. Rev. Lett.},
	author = {Hsieh, D. and Xia, Y. and Qian, D. and Wray, L. and Meier, F. and Dil, J. H. and Osterwalder, J. and Patthey, L. and Fedorov, A. V. and Lin, H. and Bansil, A. and Grauer, D. and Hor, Y. S. and Cava, R. J. and Hasan, M. Z.},
	month = sep,
	year = {2009},
	pages = {146401},
}

@article{liu_origins_2016,
	title = {Origins of enhanced thermoelectric power factor in topologically insulating {Bi$_{0.64}$Sb$_{1.36}$Te$_{3}$} thin films},
	volume = {108},
	issn = {0003-6951, 1077-3118},
	url = {https://pubs.aip.org/apl/article/108/4/043902/31583/Origins-of-enhanced-thermoelectric-power-factor-in},
	doi = {10.1063/1.4940923},
	number = {4},
	urldate = {2024-04-11},
	journal = {Appl. Phys. Lett.},
	author = {Liu, Wei and Chi, Hang and Walrath, J. C. and Chang, A. S. and Stoica, Vladimir A. and Endicott, Lynn and Tang, Xinfeng and Goldman, R. S. and Uher, Ctirad},
	month = jan,
	year = {2016},
	pages = {43902},
}

@article{liu_high-quality_2015,
	title = {High-quality ultra-flat {BiSbTe$_{3}$} films grown by {MBE}},
	volume = {410},
	issn = {00220248},
	url = {https://linkinghub.elsevier.com/retrieve/pii/S0022024814006952},
	doi = {10.1016/j.jcrysgro.2014.10.011},
	urldate = {2024-04-11},
	journal = {J. Cryst. Growth},
	author = {Liu, Wei and Endicott, Lynn and Stoica, Vladimir A. and Chi, Hang and Clarke, Roy and Uher, Ctirad},
	month = jan,
	year = {2015},
	pages = {23},
}

@article{zhang_band_2011,
	title = {Band structure engineering in {(Bi$_{1-x}$Sb$_{x}$)$_{2}$Te$_{3}$} ternary topological insulators},
	volume = {2},
	issn = {2041-1723},
	url = {https://www.nature.com/articles/ncomms1588},
	doi = {10.1038/ncomms1588},
	number = {1},
	urldate = {2024-04-11},
	journal = {Nat. Commun.},
	author = {Zhang, Jinsong and Chang, Cui-Zu and Zhang, Zuocheng and Wen, Jing and Feng, Xiao and Li, Kang and Liu, Minhao and He, Ke and Wang, Lili and Chen, Xi and Xue, Qi-Kun and Ma, Xucun and Wang, Yayu},
	month = dec,
	year = {2011},
	pages = {574},
}

@article{moore_birth_2010,
	title = {The birth of topological insulators},
	volume = {464},
	copyright = {http://www.springer.com/tdm},
	issn = {0028-0836, 1476-4687},
	url = {https://www.nature.com/articles/nature08916},
	doi = {10.1038/nature08916},
	number = {7286},
	urldate = {2024-04-11},
	journal = {Nat.},
	author = {Moore, Joel E.},
	month = mar,
	year = {2010},
	pages = {194},
}

@article{roushan_topological_2009,
	title = {Topological surface states protected from backscattering by chiral spin texture},
	volume = {460},
	copyright = {http://www.springer.com/tdm},
	issn = {0028-0836, 1476-4687},
	url = {https://www.nature.com/articles/nature08308},
	doi = {10.1038/nature08308},
	number = {7259},
	urldate = {2024-04-11},
	journal = {Nat.},
	author = {Roushan, Pedram and Seo, Jungpil and Parker, Colin V. and Hor, Y. S. and Hsieh, D. and Qian, Dong and Richardella, Anthony and Hasan, M. Z. and Cava, R. J. and Yazdani, Ali},
	month = aug,
	year = {2009},
	pages = {1106},
}

@article{hsieh_tunable_2009,
	title = {A tunable topological insulator in the spin helical {Dirac} transport regime},
	volume = {460},
	copyright = {http://www.springer.com/tdm},
	issn = {0028-0836, 1476-4687},
	url = {https://www.nature.com/articles/nature08234},
	doi = {10.1038/nature08234},
	number = {7259},
	urldate = {2024-04-11},
	journal = {Nat.},
	author = {Hsieh, D. and Xia, Y. and Qian, D. and Wray, L. and Dil, J. H. and Meier, F. and Osterwalder, J. and Patthey, L. and Checkelsky, J. G. and Ong, N. P. and Fedorov, A. V. and Lin, H. and Bansil, A. and Grauer, D. and Hor, Y. S. and Cava, R. J. and Hasan, M. Z.},
	month = aug,
	year = {2009},
	pages = {1101},
}

@article{xia_observation_2009,
	title = {Observation of a large-gap topological-insulator class with a single {Dirac} cone on the surface},
	volume = {5},
	copyright = {http://www.springer.com/tdm},
	issn = {1745-2473, 1745-2481},
	url = {https://www.nature.com/articles/nphys1274},
	doi = {10.1038/nphys1274},
	number = {6},
	urldate = {2024-04-11},
	journal = {Nat. Phys.},
	author = {Xia, Y. and Qian, D. and Hsieh, D. and Wray, L. and Pal, A. and Lin, H. and Bansil, A. and Grauer, D. and Hor, Y. S. and Cava, R. J. and Hasan, M. Z.},
	month = jun,
	year = {2009},
	pages = {398},
}

@article{zhang_topological_2009,
	title = {Topological insulators in {Bi$_{2}$Se$_{3}$}, {Bi$_{2}$Te$_{3}$} and {Sb$_{2}$Te$_{3}$} with a single {Dirac} cone on the surface},
	volume = {5},
	copyright = {http://www.springer.com/tdm},
	issn = {1745-2473, 1745-2481},
	url = {https://www.nature.com/articles/nphys1270},
	doi = {10.1038/nphys1270},
	number = {6},
	urldate = {2024-04-11},
	journal = {Nat. Phys.},
	author = {Zhang, Haijun and Liu, Chao-Xing and Qi, Xiao-Liang and Dai, Xi and Fang, Zhong and Zhang, Shou-Cheng},
	month = jun,
	year = {2009},
	pages = {438},
}

@article{chen_experimental_2009,
	title = {Experimental Realization of a Three-Dimensional Topological Insulator, {Bi}$_{\textrm{2}}${Te}$_{\textrm{3}}$},
	volume = {325},
	issn = {0036-8075, 1095-9203},
	url = {https://www.science.org/doi/10.1126/science.1173034},
	doi = {10.1126/science.1173034},
	number = {5937},
	urldate = {2024-04-11},
	journal = {Sci.},
	author = {Chen, Y. L. and Analytis, J. G. and Chu, J.-H. and Liu, Z. K. and Mo, S.-K. and Qi, X. L. and Zhang, H. J. and Lu, D. H. and Dai, X. and Fang, Z. and Zhang, S. C. and Fisher, I. R. and Hussain, Z. and Shen, Z.-X.},
	month = jul,
	year = {2009},
	pages = {178},
}

@article{kongAmbipolarFieldEffect2011,
	title = {Ambipolar field effect in the ternary topological insulator {(Bi$_{1-x}$Sb$_{x}$)$_{2}$Te$_{3}$} by composition tuning},
	volume = {6},
	copyright = {http://www.springer.com/tdm},
	issn = {1748-3387, 1748-3395},
	url = {https://www.nature.com/articles/nnano.2011.172},
	doi = {10.1038/nnano.2011.172},
	number = {11},
	urldate = {2024-10-01},
	journal = {Nat. Nanotechnol.},
	author = {Kong, Desheng and Chen, Yulin and Cha, Judy J. and Zhang, Qianfan and Analytis, James G. and Lai, Keji and Liu, Zhongkai and Hong, Seung Sae and Koski, Kristie J. and Mo, Sung-Kwan and Hussain, Zahid and Fisher, Ian R. and Shen, Zhi-Xun and Cui, Yi},
	month = nov,
	year = {2011},
	pages = {705},
}

@article{suSpinChargeConversionManipulated2021,
	title = {Spin-to-Charge Conversion Manipulated by Fine-Tuning the Fermi Level of Topological Insulator {(Bi$_{1-x}$Sb$_{x}$)$_{2}$Te$_{3}$}},
	volume = {3},
	copyright = {https://doi.org/10.15223/policy-029},
	issn = {2637-6113, 2637-6113},
	url = {https://pubs.acs.org/doi/10.1021/acsaelm.1c00182},
	doi = {10.1021/acsaelm.1c00182},
	number = {7},
	urldate = {2024-10-01},
	journal = {ACS Appl. Electron. Mater.},
	author = {Su, Shu Hsuan and Chuang, Pei-Yu and Lee, Jung-Chuan and Chong, Cheong-Wei and Li, Ya Wen and Lin, Zong Mou and Chen, Yi-Chun and Cheng, Cheng-Maw and Huang, Jung-Chun-Andrew},
	month = jul,
	year = {2021},
	pages = {2988},
}

@article{zhaoExcellentThermoelectricPerformance2021,
	title = {Excellent Thermoelectric Performance from In Situ Reaction between Co Nanoparticles and {BiSbTe} Flexible Films},
	volume = {13},
	issn = {1944-8244},
	url = {https://doi.org/10.1021/acsami.1c19143},
	doi = {10.1021/acsami.1c19143},
	number = {49},
	urldate = {2024-09-26},
	journal = {ACS Appl. Mater. Interfaces},
	author = {Zhao, Yao and Nie, Xiaolei and Sun, Congli and Chen, Yifan and Ke, Shaoqiu and Li, Cuncheng and Zhu, Wanting and Sang, Xiahan and Zhao, Wenyu and Zhang, Qingjie},
	month = dec,
	year = {2021},
	pages = {58746},
}

@article{hikami_spin-orbit_1980,
	title = {Spin-Orbit Interaction and Magnetoresistance in the Two Dimensional Random System},
	volume = {63},
	issn = {0033-068X, 1347-4081},
	url = {https://academic.oup.com/ptp/article-lookup/doi/10.1143/PTP.63.707},
	doi = {10.1143/PTP.63.707},
	number = {2},
	urldate = {2024-07-15},
	journal = {Prog. Theor. Phys.},
	author = {Hikami, S. and Larkin, A. I. and Nagaoka, Y.},
	month = feb,
	year = {1980},
	pages = {707},
}

@article{bergmann_weak_1982,
	title = {Weak anti-localization—{An} experimental proof for the destructive interference of rotated spin 1/2},
	volume = {42},
	issn = {0038-1098},
	url = {https://www.sciencedirect.com/science/article/pii/0038109882900138},
	doi = {10.1016/0038-1098(82)90013-8},
	number = {11},
	urldate = {2024-08-02},
	journal = {Solid State Commun.},
	author = {Bergmann, G.},
	month = jun,
	year = {1982},
	pages = {815},
}

@article{QE_2009,
  title={{QUANTUM} {ESPRESSO}: a modular and open-source software project for quantum simulations of materials},
  author={Giannozzi, Paolo and Baroni, Stefano and Bonini, Nicola and Calandra, Matteo and Car, Roberto and Cavazzoni, Carlo and Ceresoli, Davide and Chiarotti, Guido L and Cococcioni, Matteo and Dabo, Ismaila and Dal Corso, Andrea and De Gironcoli, Stefano and Fabris, Stefano and Fratesi, Guido and Gebauer, Ralph and Gerstmann, Uwe and Gougoussis, Christos and Kokalj, Anton and Lazzeri, Michele and Martin-Samos, Layla and Marzari, Nicola and Mauri, Francesco and Mazzarello, Riccardo and Paolini, Stefano and Pasquarello, Alfredo and Paulatto, Lorenzo and Sbraccia, Carlo and Scandolo, Sandro and Sclauzero, Gabriele and Seitsonen, Ari P and Smogunov, Alexander and Umari, Paolo and Wentzcovitch, Renata M},
  journal={J. Condens. Matter Phys.},
  volume={21},
  number={39},
  pages={395502},
  year={2009},
}

@article{QE_2017,
  title={Advanced capabilities for materials modelling with {QUANTUM} {ESPRESSO}},
  author={Giannozzi, P and Andreussi, O and Brumme, T and Bunau, O and Buongiorno Nardelli, M and Calandra, M and Car, R and Cavazzoni, C and Ceresoli, D and Cococcioni, M and Colonna, N and Carnimeo, I and Dal Corso, A and De Gironcoli, S and Delugas, P and DiStasio, R A and Ferretti, A and Floris, A and Fratesi, G and Fugallo, G and Gebauer, R and Gerstmann, U and Giustino, F and Gorni, T and Jia, J and Kawamura, M and Ko, H-Y and Kokalj, A and Küçükbenli, E and Lazzeri, M and Marsili, M and Marzari, N and Mauri, F and Nguyen, N L and Nguyen, H-V and Otero-de-la-Roza, A and Paulatto, L and Poncé, S and Rocca, D and Sabatini, R and Santra, B and Schlipf, M and Seitsonen, A P and Smogunov, A and Timrov, I and Thonhauser, T and Umari, P and Vast, N and Wu, X and Baroni, S},
  journal={J. Condens. Matter Phys.},
  volume={29},
  number={46},
  pages={465901},
  year={2017},
}

@article{chaWeakAntilocalizationBi2SexTe1x32012,
	title = {Weak Antilocalization in {Bi$_{2}$(Se$_{x}$Te$_{1-x}$)$_{3}$} Nanoribbons and Nanoplates},
	volume = {12},
	issn = {1530-6984},
	url = {https://doi.org/10.1021/nl300018j},
	doi = {10.1021/nl300018j},
	number = {2},
	urldate = {2024-10-01},
	journal = {Nano Lett.},
	author = {Cha, Judy J. and Kong, Desheng and Hong, Seung-Sae and Analytis, James G. and Lai, Keji and Cui, Yi},
	month = feb,
	year = {2012},
	pages = {1107},
}

@article{urkudeTemperatureImpurityEffect2017,
	title = {Temperature and impurity effect on parallel field magnetoconductance of bulk insulating topological insulator {(Bi$_{1-x}$Sb$_{x}$)$_{2}$Te$_{3}$}},
	volume = {29},
	issn = {0953-8984, 1361-648X},
	url = {https://iopscience.iop.org/article/10.1088/1361-648X/aa9648},
	doi = {10.1088/1361-648X/aa9648},
	number = {49},
	urldate = {2024-10-02},
	journal = {J. Phys.: Condens. Matter},
	author = {Urkude, R R and Rawat, R and Palikundwar, U A},
	month = dec,
	year = {2017},
	pages = {495602},
}

@article{steinbergElectricallyTunableSurfacebulk2011,
	title = {Electrically tunable surface-to-bulk coherent coupling in topological insulator thin films},
	volume = {84},
	copyright = {http://link.aps.org/licenses/aps-default-license},
	issn = {1098-0121, 1550-235X},
	url = {https://link.aps.org/doi/10.1103/PhysRevB.84.233101},
	doi = {10.1103/PhysRevB.84.233101},
	number = {23},
	urldate = {2024-09-27},
	journal = {Phys. Rev. B},
	author = {Steinberg, H. and Laloë, J.-B. and Fatemi, V. and Moodera, J. S. and Jarillo-Herrero, P.},
	month = dec,
	year = {2011},
	pages = {233101},
}

@article{weyrichGrowthCharacterizationTransport2016,
	title = {Growth, characterization, and transport properties of ternary {(Bi$_{1-x}$Sb$_{x}$)$_{2}$Te$_{3}$} topological insulator layers},
	volume = {28},
	issn = {0953-8984, 1361-648X},
	url = {https://iopscience.iop.org/article/10.1088/0953-8984/28/49/495501},
	doi = {10.1088/0953-8984/28/49/495501},
	number = {49},
	urldate = {2024-10-02},
	journal = {J. Phys.: Condens. Matter},
	author = {Weyrich, C and Drögeler, M and Kampmeier, J and Eschbach, M and Mussler, G and Merzenich, T and Stoica, T and Batov, I E and Schubert, J and Plucinski, L and Beschoten, B and Schneider, C M and Stampfer, C and Grützmacher, D and Schäpers, Th},
	month = dec,
	year = {2016},
	pages = {495501},
}

@article{leThicknessdependentMagnetotransportProperties2017,
	title = {Thickness-dependent magnetotransport properties and terahertz response of topological insulator {Bi$_{2}$Te$_{3}$} thin films},
	volume = {692},
	issn = {09258388},
	url = {https://linkinghub.elsevier.com/retrieve/pii/S0925838816328614},
	doi = {10.1016/j.jallcom.2016.09.109},
	urldate = {2024-09-27},
	journal = {J. Alloys Compd.},
	author = {Le, Phuoc Huu and Liu, Po-Tsun and Luo, Chih Wei and Lin, Jiunn-Yuan and Wu, Kaung Hsiung},
	month = jan,
	year = {2017},
	pages = {972},
}

@article{matz_non-ohmic_1964,
	title = {Non-Ohmic Transport in Semiconductors in a Magnetic Field},
	volume = {5},
	issn = {1521-3951},
	doi = {10.1002/pssb.19640050306},
	number = {3},
	journal = {Phys. Status Solidi B},
	author = {Matz, D. and Garcia-Moliner, F.},
	year = {1964},
	pages = {495},
}

@article{hattori_note_1960,
	title = {Note on the Field Dependence of the Mobility in Semiconductors},
	volume = {15},
	doi = {10.1143/JPSJ.15.1237},
	number = {7},
	journal = {J. Phys. Soc. Jpn.},
	author = {Hattori, Masumi and Sato, Hisanao},
	month = jul,
	year = {1960},
	pages = {1237},
}

@article{liao_enhanced_2017,
	title = {Enhanced electron dephasing in three-dimensional topological insulators},
	volume = {8},
	issn = {2041-1723},
	doi = {10.1038/ncomms16071},
	number = {1},
	urldate = {2024-07-28},
	journal = {Nat. Commun.},
	author = {Liao, Jian and Ou, Yunbo and Liu, Haiwen and He, Ke and Ma, Xucun and Xue, Qi-Kun and Li, Yongqing},
	month = jul,
	year = {2017},
	pages = {16071},
}

@article{beidenkopf_spatial_2011,
	title = {Spatial fluctuations of helical {Dirac} fermions on the surface of topological insulators},
	volume = {7},
	issn = {1745-2481},
	doi = {10.1038/nphys2108},
	number = {12},
	journal = {Nat. Phys.},
	author = {Beidenkopf, Haim and Roushan, Pedram and Seo, Jungpil and Gorman, Lindsay and Drozdov, Ilya and Hor, Yew San and Cava, R. J. and Yazdani, Ali},
	year = {2011},
	pages = {939},
}

@article{yuRobustGaplessSurface2020,
	title = {Robust Gapless Surface State against Surface Magnetic Impurities on {Bi$_{0.5}$Sb$_{0.5}$Te$_{3}$} Evidenced by \textit{{in} {situ}} Magnetotransport Measurements},
	volume = {124},
	issn = {0031-9007, 1079-7114},
	url = {https://link.aps.org/doi/10.1103/PhysRevLett.124.126601},
	doi = {10.1103/PhysRevLett.124.126601},
	number = {12},
	urldate = {2024-10-02},
	journal = {Phys. Rev. Lett.},
	author = {Yu, Liuqi and Hu, Longqian and Barreda, Jorge L. and Guan, Tong and He, Xiaoyue and Wu, Kehui and Li, Yongqing and Xiong, Peng},
	year = {2020},
	pages = {126601},
}

@article{kimFreeelectronCreation60deg2016,
	title = {Free-electron creation at the 60° twin boundary in {Bi$_{2}$Te$_{3}$}},
	volume = {7},
	issn = {2041-1723},
	url = {https://www.nature.com/articles/ncomms12449},
	doi = {10.1038/ncomms12449},
	number = {1},
	urldate = {2024-09-19},
	journal = {Nat Commun.},
	author = {Kim, Kwang-Chon and Lee, Joohwi and Kim, Byung Kyu and Choi, Won Young and Chang, Hye Jung and Won, Sung Ok and Kwon, Beomjin and Kim, Seong Keun and Hyun, Dow-Bin and Kim, Hyun Jae and Koo, Hyun Cheol and Choi, Jung-Hae and Kim, Dong-Ik and Kim, Jin-Sang and Baek, Seung-Hyub},
	year = {2016},
	pages = {12449},
}

@article{shaoEpitaxialGrowthFlat2018,
	title = {Epitaxial Growth of Flat Antimonene Monolayer: A New Honeycomb Analogue of Graphene},
	volume = {18},
	issn = {1530-6984},
	url = {https://doi.org/10.1021/acs.nanolett.8b00429},
	doi = {10.1021/acs.nanolett.8b00429},
	number = {3},
	urldate = {2024-11-06},
	journal = {Nano Lett.},
	author = {Shao, Yan and Liu, Zhong-Liu and Cheng, Cai and Wu, Xu and Liu, Hang and Liu, Chen and Wang, Jia-Ou and Zhu, Shi-Yu and Wang, Yu-Qi and Shi, Dong-Xia and Ibrahim, Kurash and Sun, Jia-Tao and Wang, Ye-Liang and Gao, Hong-Jun},
	month = mar,
	year = {2018},
	pages = {2133},
}

@article{mazumderBriefReviewBi2Se32021,
	title = {A brief review of {Bi$_{2}$Se$_{3}$} based topological insulator: {From} fundamentals to applications},
	volume = {888},
	issn = {09258388},
	shorttitle = {A brief review of {Bi2Se3} based topological insulator},
	url = {https://linkinghub.elsevier.com/retrieve/pii/S0925838821029017},
	doi = {10.1016/j.jallcom.2021.161492},
	urldate = {2025-02-18},
	journal = {J. Alloys Compd.},
	author = {Mazumder, Kushal and Shirage, Parasharam M.},
	year = {2021},
	pages = {161492},
}

@article{heremansTetradymitesThermoelectricsTopological2017a,
	title = {Tetradymites as thermoelectrics and topological insulators},
	volume = {2},
	issn = {2058-8437},
	url = {https://www.nature.com/articles/natrevmats201749},
	doi = {10.1038/natrevmats.2017.49},
	number = {10},
	urldate = {2024-10-02},
	journal = {Nat. Rev. Mater.},
	author = {Heremans, Joseph P. and Cava, Robert J. and Samarth, Nitin},
	month = sep,
	year = {2017},
	pages = {17049},
}

@article{tangComprehensiveReviewBi2Te3based2022,
	title = {A comprehensive review on {Bi$_{2}$Te$_{3}$}-based thin films: {Thermoelectrics} and beyond},
	volume = {1},
	copyright = {© 2022 The Authors. Interdisciplinary Materials published by Wuhan University of Technology and John Wiley \& Sons Australia, Ltd.},
	issn = {2767-441X},
	shorttitle = {A comprehensive review on {Bi2Te3}-based thin films},
	url = {https://onlinelibrary.wiley.com/doi/abs/10.1002/idm2.12009},
	doi = {10.1002/idm2.12009},
	number = {1},
	urldate = {2025-02-19},
	journal = {Interd. Mater.},
	author = {Tang, Xinfeng and Li, Ziwei and Liu, Wei and Zhang, Qingjie and Uher, Ctirad},
	year = {2022},
	pages = {88},

}

@article{ranOnedimensionalTopologicallyProtected2009,
	title = {One-dimensional topologically protected modes in topological insulators with lattice dislocations},
	volume = {5},
	copyright = {http://www.springer.com/tdm},
	issn = {1745-2473, 1745-2481},
	url = {https://www.nature.com/articles/nphys1220},
	doi = {10.1038/nphys1220},
	number = {4},
	urldate = {2025-02-25},
	journal = {Nat. Phys.},
	author = {Ran, Ying and Zhang, Yi and Vishwanath, Ashvin},
	month = apr,
	year = {2009},
	pages = {298},
}

@article{liu_tuning_2014,
	title = {Tuning {Dirac} states by strain in the topological insulator {Bi$_2$Se$_3$}},
	volume = {10},
	doi = {10.1038/nphys2898},
	number = {4},
	journal = {Nat. Phys.},
	author = {Liu, Y. and Li, Y. Y. and Rajput, S. and Gilks, D. and Lari, L. and Galindo, P. L. and Weinert, M. and Lazarov, V. K. and Li, L.},
	month = apr,
	year = {2014},
	pages = {294--299},
}

\end{document}


\nolinenumbers 

\title{\textbf{Supplementary Material}\\
\textbf{Probing Topological Surface States and Conduction via Extended Defects in (Bi$_{1-x}$Sb$_{x}$)$_{2}$Te$_{3}$ Films}
}%

\author{A. Liu}
\affiliation{Department of Materials Science and Engineering, University of Michigan, Ann Arbor, MI 48109, USA}
\author{A. Gil}
\affiliation{Applied Physics Program, University of Michigan, Ann Arbor, MI 48109, USA}
\author{M. Choi}
\affiliation{Department of Mechanical Science and Engineering, University of Illinois at Urbana-Champaign, Champaign, IL 61801, USA}
\affiliation{Materials Research Laboratory, University of Illinois Urbana-Champaign, Urbana, Illinois, 61801, USA}
\author{B. Akgenc Hanedar}
\affiliation{Department of Physics, Kirklareli University, Kirklareli 39100, Türkiye}
\author{V. Stoica}
\affiliation{Department of Physics, University of Michigan, Ann Arbor, MI 48109, USA}
\affiliation{Department of Materials Science and Engineering, Penn State University, University Park, PA, 16802, USA}
\author{Z. You}
\affiliation{Department of Physics, University of Michigan, Ann Arbor, MI 48109, USA}
\author{S. Sinha}
\affiliation{Applied Physics Program, University of Michigan, Ann Arbor, MI 48109, USA}
\author{T. Hakioglu}
\affiliation{Department of Physics, University of Michigan, Ann Arbor, MI 48109, USA}
\affiliation{Energy Institute, Istanbul Technical University, Maslak, Sarıyer, 34469, Istanbul, Türkiye}
\author{H. T. Johnson}
\affiliation{Department of Mechanical Science and Engineering, University of Illinois at Urbana-Champaign, Champaign, IL 61801, USA}
\affiliation{Materials Research Laboratory, University of Illinois Urbana-Champaign, Urbana, Illinois, 61801, USA}
\author{K. Sun}
\affiliation{Department of Physics, University of Michigan, Ann Arbor, MI 48109, USA}
\author{R. Clarke}
\affiliation{Applied Physics Program, University of Michigan, Ann Arbor, MI 48109, USA}
\affiliation{Department of Physics, University of Michigan, Ann Arbor, MI 48109, USA}
\author{C. Uher}
\affiliation{Department of Physics, University of Michigan, Ann Arbor, MI 48109, USA}
\author{C. Kurdak}
\affiliation{Applied Physics Program, University of Michigan, Ann Arbor, MI 48109, USA}
\affiliation{Department of Physics, University of Michigan, Ann Arbor, MI 48109, USA}
\author{R. S. Goldman}
 \thanks{Corresponding author: rsgold@umich.edu}
\affiliation{Department of Materials Science and Engineering, University of Michigan, Ann Arbor, MI 48109, USA}
\affiliation{Applied Physics Program, University of Michigan, Ann Arbor, MI 48109, USA}
\affiliation{Department of Physics, University of Michigan, Ann Arbor, MI 48109, USA}


\begin{abstract}

We present experimental and computational methods, including Hall bar contact misalignment corrections, analysis of scanning tunneling spectroscopy (STS) data, density of states (DOS) calculations, and tight-binding model calculations of twin boundary modes.
\end{abstract}

\maketitle
\clearpage



\section{\label{sec:transport-1_Supp}Hall Bar Contact Misalignment Corrections}

\begin{figure}[H]
    \centering
    \includegraphics[width=\linewidth]{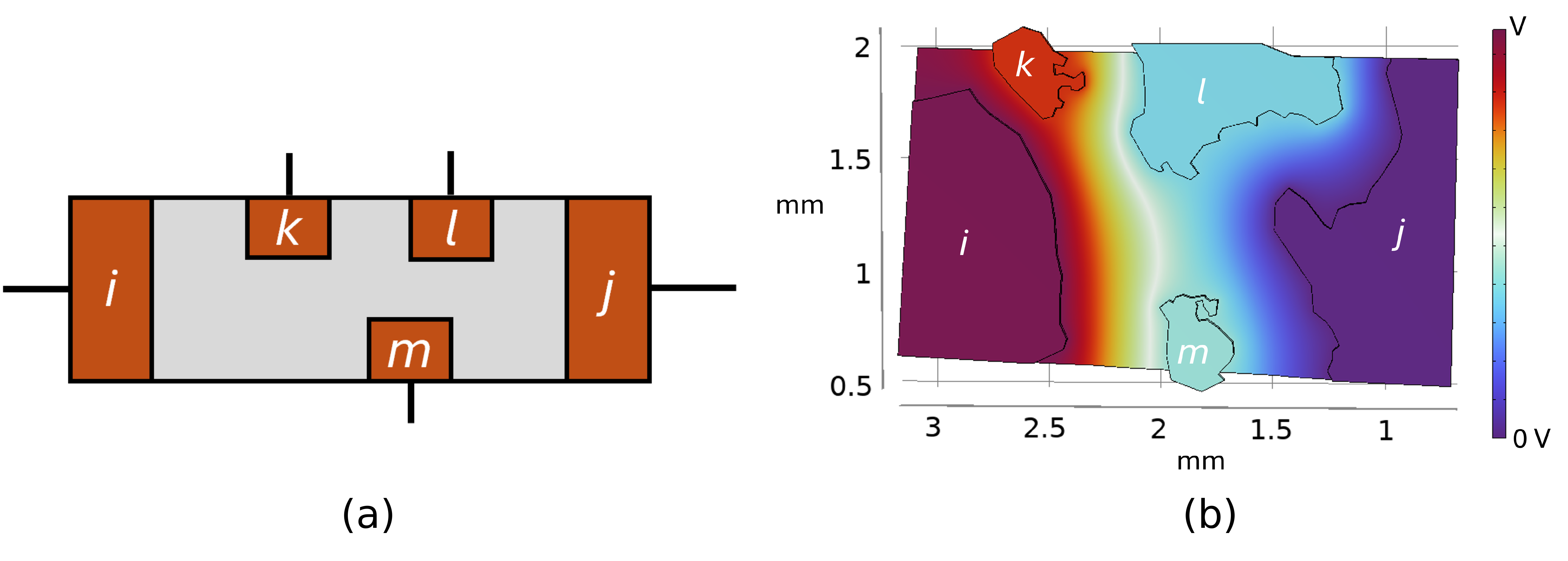}
        \caption{(a): Illustration of Hall bar utilized for magnetotransport measurements, with corresponding (b) COMSOL simulations of the electric potential profile for large, irregularly-shaped contacts. In (a), indium contacts are represented by orange boxes. $I_{i,j}$ flows between contacts $i$ and $j$ while longitudinal (lateral) voltage is measured through $k$ and $l$ ($m$ and $l$) in a four-terminal configuration. In (b), contacts are labeled equivalently. Currents and voltages are simulated to calculate conversion factors $\alpha$ and $\beta$.}
        \label{fig: mt_schem_com}
\end{figure}

A typical schematic of a Hall bar used for magnetotransport measurements is shown in Fig.\ \ref{fig: mt_schem_com}(a). A current $I_{i,j}$ is passed between contacts $i$ and $j$, with sufficiently low bias currents to ensure that voltage measurements are independent of the bias current. Longitudinal and lateral voltages, $V_{k,l}$ and $V_{m,l}$, were measured simultaneously as a function of magnetic field and temperature using standard lock-in techniques. Since the indium contacts were hand-soldered onto the samples, contact misalignment and oversizing need to be considered in the data analysis, especially when converting raw voltages to resistance and conductance.

First, voltage data are converted to resistance via Ohm's Law; for example, the longitudinal resistance is calculated as $R_{i,j;k,l} = V_{k,l}/I_{i,j}$. Contact misalignment is addressed by symmetrizing the longitudinal resistance and anti-symmetrizing the Hall resistance:
\begin{subequations}
    \begin{align}
        R_{i,j;k,l}^{sym} & = \frac{1}{2}\big[R_{i,j;k,l}(B) + R_{i,j;k,l}(-B)\big]\\
        R_{i,j;m,l}^{asym} & = \frac{1}{2}\big[R_{i,j;m,l}(B) - R_{i,j;m,l}(-B)\big].
    \end{align}
\end{subequations}

To convert raw resistances to sheet and Hall resistance, we use the relations:
\begin{subequations}
  \begin{align} 
        R_{i,j;k,l} & = \alpha R_{sheet}\\
        R_{i,j;m,l} & = \beta R_{\text{Hall}}
  \end{align}
  \label{eqn: mt conv factors}
\end{subequations}
\noindent where $\alpha$ and $\beta$ are sample-geometry-dependent factors determined using COMSOL finite element analysis simulations. For example, a COMSOL simulation of irregularly-shaped contacts on a Hall bar are shown in Fig.\ \ref{fig: mt_schem_com}(b) with $R_{sheet}\approx$ 1 k$\Omega$. An arbitrary current $I_{i,j}$ is simulated between contacts $i$ and $j$; Voltages $V_{k,l}$ and $V_{m,l}$ are then computed and converted to resistance, yielding $R_{i,j;k,l}$ and $R_{i,j;m,l}$. Finally, using Eq.\ \eqref{eqn: mt conv factors}, $\alpha$ and $\beta$ are calculated and applied to the symmetrized and anti-symmetrized measured resistances, thus providing $R_{sheet}$ and $R_{\text{Hall}}$. Finally,  a matrix inversion of the resistance values was used for matrix inversion of the resistance values, as follows:
\begin{subequations}
    \begin{align} 
        \sigma_{xx} = \frac{R_{sheet}}{R_{sheet}^2 + R_{\text{Hall}}^2} \label{eqn: sigxx data}\\
        \sigma_{xy} = \frac{R_{\text{Hall}}}{R_{sheet}^2 + R_{\text{Hall}}^2}.
    \end{align}
    \label{eqn: mt conductance}
\end{subequations}

\section{\label{sec:STM}Analysis of Scanning Tunneling Spectroscopy Data} 

The scanning tunneling spectroscopy (STS) data were smoothed using MATLAB’s ‘smooth’ function, which applies an adjacent-averaging algorithm. A smoothing constant was selected to ensure that the conductance values were averaged over an energy scale smaller than $k_B T$. The data were then normalized to a range of 0 to 1. A boundary condition was set such that $dI/dV$ = 0 corresponds to 0, and a range for the valence and conduction bands was defined, within which linear least-squares fits were performed using MATLAB’s ‘polyfit’ function. The edges of the valence and conduction bands were determined by the intercepts of these linear fits with $dI/dV$ = 0. The effective band gap was calculated as the energy difference between the conduction and valence band edges.


\section{\label{sec:DOS calculation}Density of States Calculations}
Density of states (DOS) calculations were performed using QUANTUM ESPRESSO \cite{QE_2009,QE_2017} with the Projector Augmented Wave (PAW) method to describe electron interactions. In the PAW pseudopotential, We consider 15 electrons for Bi ($5d^{10}6s^26p^3$), 5 electrons for Sb ($5s^25p^3$), and 6 electrons for Te ($5s^25p^4$). For the wave functions, the plane-wave cutoff energy is set to 680 eV. We adopt the Fermi-Dirac smearing method with a smearing width of 0.001 eV. Spin-orbit coupling effects are included for all structures: bulk, basal twin, and 60$^{\circ}$ twin, which contain 30, 30, and 120 atoms, respectively. Brillouin zone integrations are carried out using a Monkhorst-Pack k-point mesh: a $12 \times 12 \times 2$ grid for both the bulk and basal twin structures, and a $2 \times 12 \times 2$ grid for the $60^{\circ}$ twin structure, to ensure accurate convergence of the DOS.
\begin{figure}[h]
    \centering
    \includegraphics[width=0.9\linewidth]{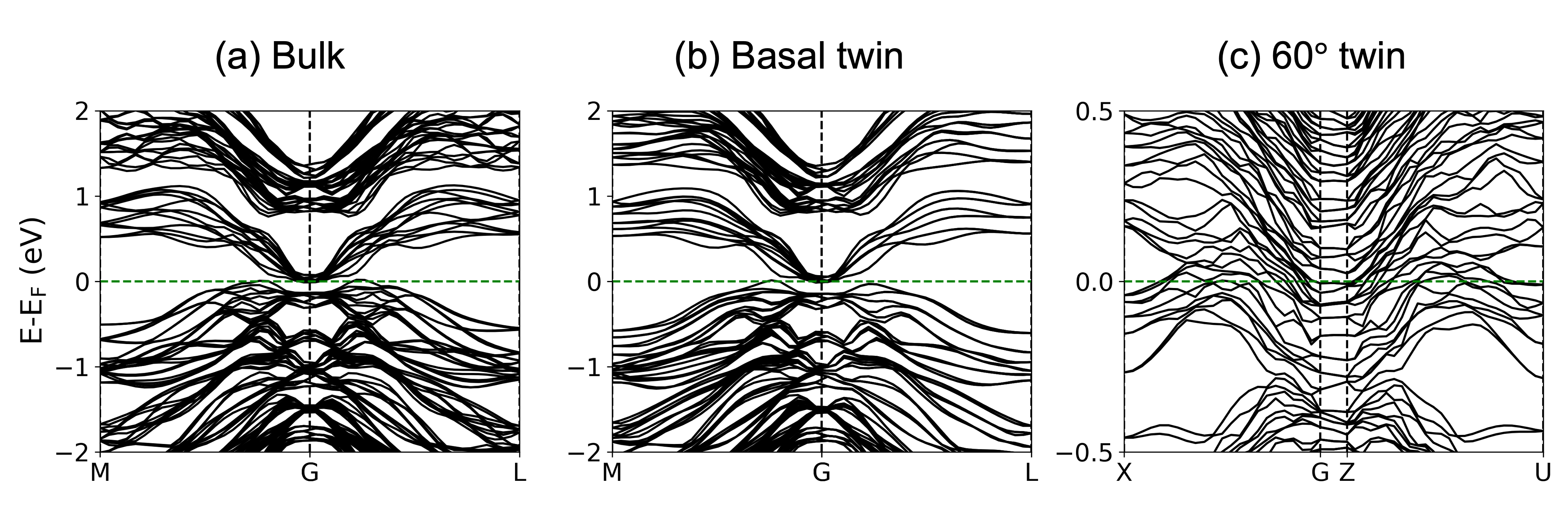}
    \caption{Band structures of bulk, basal twin, and 60$\degree$ twin structures.}
    \label{fig:band_structure_twin}
\end{figure}
Fig.\ \ref{fig:band_structure_twin} shows the band structures of bulk, basal twin, and 60$^\circ$ twin structures. Note that the set of special points ($MGL$) in the bulk and basal twin structures is different from the set ($XGZU$) in 60$^{\circ}$ twin structure due to the difference in symmetry between hexagonal and orthorhombic unit cells. Fig.\ \ref{fig:band_structure_twin} (c) shows a high density of electron bands near the Fermi energy, which results in a high DOS near the Fermi energy.


\section{\label{sec: Origin of Twin-Boundary Modes}Tight Binding Calculations of Twin-Boundary Modes}

For these calculations, we consider a cubic-lattice version of the Bernevig-Hughes-Zhang model~\cite{bernevig_BHZ_model2006}. Each unit cell of this cubic lattice contains two orbitals, each with two-fold spin degeneracy. These two orbitals have opposite parity under spatial inversion, with one being even and the other odd. The resulting two energy bands formed by these two orbitals exhibit a band inversion. For simplicity, we set the band inversion to occur at the $\Gamma$ point of the bulk Brillouin zone and focus on the twin boundary along the (1,0,0) crystal planes. However, it is important to note that the conclusions discussed below are insensitive to these specific details and remain qualitatively robust for any topological insulator and any orientation of the twin boundaries.

In momentum space, the model can be represented as a $4\times 4$ matrix, since each unit cell contains four quantum states: two orbitals with two-fold spin degeneracy. The Hamiltonian is given by
\begin{align}
H=
\begin{pmatrix}
\epsilon_e(\mathbf{k}) \; \mathbf{I} & 2\Delta \sin \mathbf{k}\cdot \boldsymbol{\sigma}
\\
2\Delta \sin \mathbf{k}\cdot \boldsymbol{\sigma}&\epsilon_o(\mathbf{k}) \; \mathbf{I}
\end{pmatrix}
\label{eq:model}
\end{align}
where $\mathbf{I}$ is the $2\times 2$ identity matrix, $\boldsymbol{\sigma}=(\sigma_x,\sigma_y,\sigma_z)$ represents the Pauli matrices, and $\mathbf{k}=(k_x,k_y,k_z)$ is the wavevector. Here, $\epsilon_e$ and $\epsilon_o$ denote the dispersion relations for the even and odd orbitals, respectively, in the absence of band hybridization between these two orbitals:
\begin{align}
\epsilon_e(\mathbf{k})=-2t_e (\cos k_x+\cos k_y+\cos k_z)+u
\label{eq:model_epsilon_e}
\\
\epsilon_o(\mathbf{k})=-2t_o (\cos k_x+\cos k_y+\cos k_z)-u
\label{eq:model_epsilon_o}
\end{align}
where $t_e$ ($t_0$) represents the hopping strength between neighboring orbitals with even (odd) parity. The coefficient $\Delta$ denotes the spin-dependent hopping strength between even and odd parity orbitals in neighboring unit cells, encoding the effects of spin-orbit coupling.

To simulate a twin boundary, we consider the setup shown in the right panel of Fig. 4(a) of the manuscript, where the light blue region marks the interface between two domains. The presence of the twin boundary reduces the electron tunneling strength across it (along the dashed lines), making it weaker than the tunneling between neighboring unit cells in regions without a twin boundary (along the solid lines).
In our model calculation, we define the tunneling strengths between two neighboring sites connected by a solid line as $t_o$, $t_e$, and $\Delta$, following Eqs.~\eqref{eq:model}-\eqref{eq:model_epsilon_o}. For hoppings across the twin boundary (i.e., between sites connected by a dashed line), we introduce a reduction factor $\lambda$, setting the modified tunneling strengths to $\lambda t_o$, $\lambda t_e$, and $\lambda \Delta$. Here, $\lambda$ is a control parameter in the range $0 \leq \lambda \leq 1$, quantifying the suppression of tunneling due to the twin boundary.
For the calculations below, we choose $t_e=+1$, $t_o=-1$, $u=5$, and $\Delta=0.7$. It is important to note that the qualitative features discussed here remain robust with respect to these parameter choices, as long as the model includes a band inversion, ensuring a topologically nontrivial band structure. 

\section*{References}
\bibliography{References}